\documentclass[twocolumn,showpacs,showkeys,preprintnumbers,amsmath,amssymb,pre,floatfix]{revtex4}

\usepackage{graphicx}
\usepackage{dcolumn}
\usepackage{bm}
\usepackage{psfrag}

\begin{document}
\title{Noise-assisted quantum
electron transfer in photosynthetic complexes}

\author{Alexander I. Nesterov}
   \email{nesterov@cencar.udg.mx}
\affiliation{Departamento de F{\'\i}sica, CUCEI, Universidad de Guadalajara,
Av. Revoluci\'on 1500, Guadalajara, CP 44420, Jalisco, M\'exico}
\author{Gennady P.  Berman}
 \email{gpb@lanl.gov}
\affiliation{Theoretical Division, Los Alamos National Laboratory,
Los Alamos, NM 87544, USA}

\author{Jos\'e Manuel S\'anchez Mart\'inez}
   \email{jmsm.manuel@gmail.com}
\affiliation{Departamento de F{\'\i}sica, CUCEI, Universidad de Guadalajara,
Av. Revoluci\'on 1500, Guadalajara, CP 44420, Jalisco, M\'exico}

\author{Richard T.~Sayre}
\email{ rsayre@newmexicoconsortium.org}
\affiliation{Los Alamos National Laboratory and New Mexico Consortium, 4200 W James Rd, Los Alamos,  NM, 87544, USA}

\date{\today}

\begin{abstract}
Electron transfer (ET)  between primary electron  donors and acceptors  is modeled in the photosystem II reaction center (RC).  Our model  includes (i) two discrete energy  levels associated with  donor and acceptor, interacting through a dipole-type matrix element and (ii) two continuum manifolds of electron energy levels (``sinks"), which interact directly with the donor and  acceptor.  Namely, two discrete energy levels of the donor and acceptor are embedded in their independent sinks through the corresponding interaction matrix elements. We also introduce classical (external) noise which acts simultaneously on the donor and acceptor (collective interaction). We derive a closed system of integro-differential equations  which describes the non-Markovian quantum dynamics of  the ET. A region of parameters is found in which the ET dynamics can be simplified, and described by coupled ordinary differential equations. Using these simplified equations, both  sharp and flat redox potentials are analyzed. We analytically and numerically obtain the  characteristic parameters that optimize the ET rates and  efficiency in this system.

\end{abstract}


 \preprint{LA-UR-13-23059}
\maketitle

\section{Introduction}

In photosynthetic complexes of plants, eukaryotic algae and cyanobacteria, quanta of light excite chlorophyll dipoles in the antennas of the light harvesting complexes. These local energy excitations are then transferred to the reaction centers (RCs) of photosystem I (PSI) and photosystem II (PSII), where the charge separation occurs. During charge separation an electron jumps from the donor to the acceptor and both donor and acceptor become charged. These electron jumps continue through out the whole chain of the redox potential. The characteristic time-scales of the electron dynamics vary from a few picoseconds to milliseconds. The primary charge separation occurs on a very short time-scale, of a few picoseconds \cite{BER1,GSR, XSG,Psr,Len}. Because this time-scale is so short, even the room-temperature fluctuations of the protein environment do not destroy the quantum coherent effects, which were recently discovered in these complexes \cite{Len,ECR,CWWC,PHFC,IFG,RMKL,CFMB,PPM,PPM1}.

Generally, to describe the motion of an electron between the donor and acceptor, different approaches can be used \cite{MBS,Len,HDR,CWW,MRSN,XSK,SWB,MFL,BPM}. The leading approach is based on the well-known Markus theory, which takes into account both the strength of donor-acceptor interaction, distance and the dynamics of the protein environment \cite{HDR}. In this way, the election transfer (ET) rate and the efficiency of the ET can be calculated \cite{MBS}. At the same time, many efforts have been devoted to various modifications of the Markus theory. One of them is related to taking into account the sinks which represent the continuum electron energy reservoirs analogous to the HOMO and LUMO orbitals in the chemical systems \cite{PPM,PPM1,THA} (see also references therein). These continuum manifolds are similar to those described by the Weisskopf-Wigner model \cite{WW,SM}. These sinks serve as additions to the thermal reservoirs in  ET. The principal difference between the protein (vibrational) and sink reservoirs is that the former behave as the bosonic (electromagnetic) environments  while the later provide additional electron quasi-degenerate states. These sinks increase the entropy for an electron escaping into the sinks. As a result,  the sinks modify the form of the Gibbs equilibrium distribution. Indeed, even in the case of a flat redox potential one can expect the acceptor to be populated with high enough efficiency.

In this paper, we consider ET in a model which consists of two discrete energy states, donor and acceptor. A dipole-type matrix element of a direct donor-acceptor interaction is included. The donor and acceptor interact directly with two independent sinks which are represented by two quasi-degenerate (continuum) manifolds of the electron energy levels. In addition, classical (external) noise interacts with the donor and the acceptor. By taking into account these effects, we describe the ET dynamics in different parameter region. We obtain the conditions when the system of integro-differential equations, which describes generally the non-Markovian electron dynamics, can be reduced to much simpler system of ordinary differential equations. We obtain analytically and numerically the ET for both sharp and flat redox potentials, and for different amplitudes of noise. The advantage of our approach is that it is a rigorous one, so all approximations are controlled and justified. The simultaneous contributions of the sinks and noise are analyzed, allowing us to derive optimal ET rates for various regions of parameters.

 Our paper is organized as follows. In Section II, we describe our simplified model with a single sink interacting only with the acceptor level, and present an  analytic solution for the ET, in the absence of noise. In Section III, we analyze analytically and numerically the simultaneous influence of the single acceptor sink and noise. We derive a closed system of integro-differential equations which describe the ET. We also find the region of parameters for which a system of ordinary differential equations can be applied. In Section IV, we analyze analytically and numerically  the simultaneous influence of two sinks and noise. In Section V, we discuss the obtained results. In the Appendices some useful formulae are presented.

\section{Model description}

\begin{figure}[tbh]
\scalebox{0.35}{\includegraphics{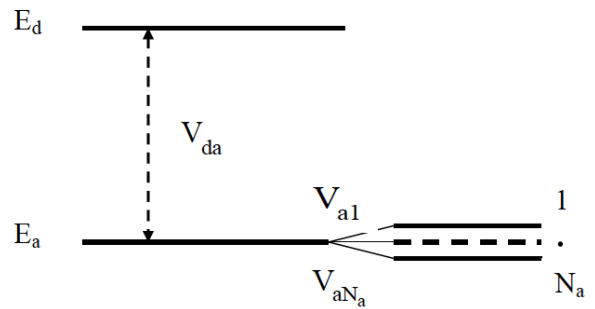}}
\caption{Schematic of our model consisting of donor and acceptor discrete energy levels, with the acceptor coupled to a sink reservoir with a nearly continuous spectrum.
\label{S1a}}
\end{figure}

Our model consists of two protein cofactors in the RC (donor and acceptor) each with a discrete energy level. In this Section, we consider, for simplicity, only a single sink which interacts with the acceptor level (the acceptor is also embedded in a sink). The sink can be considered either as an additional, third cofactor, or as a part of the acceptor. (See Fig. 1.) The first site,  denoted by $|d\rangle$, is the electron donor, with the energy level, $E_d$. The second site, $|a\rangle$, is the electron acceptor, with energy level, $E_a$.  We model the sink by a large number of discrete and nearly degenerate energy levels, $N_a \gg 1$ (Fig. \ref{S1a}). Then, the transition to the high density of states (in the continuum limit) is done for the sink.
Note that in some situations, the sinks can be considered as approximations to HOMO and LUMO orbitals with finite densities of states.

The Hamiltonian of this system can be written as
\begin{align}
H_t = E_d|d\rangle \langle d|+
E_a|a\rangle \langle a| + \frac{V}{2}(|d\rangle \langle a|+ |a\rangle \langle d|) \nonumber \\
+\sum^N_{i=1}E_i |i\rangle\langle  i | + \sum^{N_a}_{i=1} \big( V_{ai}|a\rangle\langle  i | + V_{ia}|i\rangle\langle a |\big),
\end{align}
where $E_i$ are energies of the sink levels, and $V_{da} =V/2$. (See Fig. \ref{S1a}.)

Using the standard Feshbach projection method \cite{SM,RI,RI2,RI3,VZ}, one can show that the
dynamics of the donor-acceptor (intrinsic) states can be described by the following Schr\"odinger equation with an effective non-Hermitian Hamiltonian, $ \tilde{\mathcal H}= {\mathcal H}- i \mathcal W$ (we set  $\hbar = 1$)\cite{NBB1}:
\begin{align}\label{S1b}
   i\frac{\partial \psi  (t)}{\partial t}= \tilde{\mathcal H} \psi (t),
\end{align}
where
\begin{align}
{\mathcal H} = \varepsilon_d|d\rangle \langle d|+ \varepsilon_a|a\rangle \langle a|+ \frac{V}{2}(|d\rangle \langle a|+ |a\rangle \langle d|)
\end{align}
is the dressed donor-acceptor Hamiltonian, and $\mathcal W = ({\Gamma_a}/{2})|a\rangle \langle a|$. Here  $\Gamma_a$ is the rate describing the tunneling from the acceptor to the sink. (See Appendix A for details.)

Equivalently, the dynamics of this system can be described by the Liouville  equation,
\begin{eqnarray}\label{DM1r}
   i \dot{ \rho} = [\mathcal H,\rho] - i\{\mathcal W,\rho\},
\end{eqnarray}
where   $\rho $ is the density matrix projected on the intrinsic states, and $\{\mathcal W,\rho\}= \mathcal W\rho +\rho\mathcal  W$.

The solution of the eigenvalue problem for the effective non-Hermitian Hamiltonian, $ \tilde{\mathcal H}$, yields two complex eigenvalues:  $\tilde E_{1,2}  =\tilde\lambda_{0}/2 \pm \Omega/2$,
where $\tilde\lambda_{0} =\varepsilon_d +\varepsilon_a -i\Gamma$ and $\Omega = \sqrt{V^2 +(\varepsilon +i\Gamma)^2}$.  Here we denote $\varepsilon =\varepsilon_d -\varepsilon_a$ and $\Gamma=\Gamma_a/2$.
The eigenvalues coalescence in the so-called exceptional point (EP) defined by the equation, $\Omega(V,\varepsilon,\Gamma) =0$.   Since $\Omega$ is a complex function of its parameters, we obtain two real equations: $\Re \Omega =0$ and $\Im \Omega =0$.  One can show that these equations are equivalent to $\varepsilon =0$ and $V=\Gamma$.

Note that, in contrast to the case of an Hermitian Hamiltonian, where the degeneracy is
referred to as a ``conical intersection"  (known also as
a ``diabolic point" \cite{B0}),  the coalescence of eigenvalues
results in different eigenvectors. At the EP, the eigenvectors merge,
forming a Jordan block. (See the review \cite{B}, and references therein.)

{\em Choice of parameters.} This model involves various parameters, whose values are only partially known.  Our choice of parameters is based on the data taken  for the ET through the active pathway in the quinone-type of the Photosystem II RC \cite{LVD}.  (Note that the values of parameters in energy units can be obtained by multiplying our values by $\hbar\approx 6.58\times 10^{-13}\rm meVs$. For example, $\varepsilon = 60\; \rm ps^{-1}\approx 40\rm meV$.)

\subsection{Tunneling to the sink}

In this section we discuss the ET to the sink. (For details see \cite{NBB1}.)  We assume that initially the electron occupies the upper level (donor),  $\rho_{11}(0)=1$ and $\rho_{22}(0)=0$.  With these initial conditions, the solution of the Liouville equation (\ref{DM1}) for the diagonal component of the density matrix is:
\begin{align}
\rho_{11}(t)  =& {e^{-\Gamma t}} \bigg |\Big(\cos\frac{\Omega t}{2} - i\cos\theta\sin\frac{\Omega t}{2}\Big)\bigg|^2, \\
\rho_{22}(t)  = &{e^{-\Gamma t}} \bigg |\sin\theta\sin\frac{\Omega t}{2}\bigg |^2,
\label{P3d}
\end{align}
where $\Omega= \sqrt{V^2 +(\varepsilon + i\Gamma )^2}$ is the complex Rabi frequency, $\cos\theta = (\varepsilon + i\Gamma )/\Omega$, and $\sin\theta = V/\Omega$.

Setting $\Omega= \Omega_1 +i \Omega_2 = \sqrt{p +i q}$, where
$p=  V^2+ \varepsilon^2- \Gamma^2$ and $q= 2\varepsilon\Gamma$, we obtain
\begin{align}
\Omega_1= &\pm\frac{1}{\sqrt{2}} \sqrt{p + \sqrt{p^2 + q^2}}, \\
\Omega_2= &\pm\frac{1}{\sqrt{2}} \sqrt{-p + \sqrt{p^2 + q^2}}.
\end{align}
Using these results, we obtain for $\rho_{22}(t) $ the simple analytical expression:
\begin{align}
\rho_{22}(t)  = \frac{  V^2 e^{-\Gamma t}}{2(\Omega^2_1 +\Omega^2_2)} \big (\cosh{\Omega_2 t} - \cos{\Omega_1 t} \big ).
\label{P3er}
\end{align}

We define the ET efficiency of tunneling to the sink as
\begin{align}\label{ET1}
\eta(t) = 1 - {\rm Tr}(\rho(t)) =  \int_0^t {\rm Tr}\{\mathcal W,\rho(\tau)\} d \tau.
\end{align}
This can be recast as the integrated probability of trapping the electron in the sink \cite{RMKL,CDCH},
\begin{eqnarray}
\eta(t) = 2\Gamma \int_0^t \rho_{22}(\tau)d \tau.
\label{Eq16ar}
\end{eqnarray}
Inserting $\rho_{22}(t)$ into (\ref{Eq16ar}) and performing the integration, we find that the ET efficiency is given by
\begin{align}\label{Eq17r}
 \eta(t) =& 1- \frac{ e^{-\Gamma t}}{\Gamma(\Omega^2_1 +\Omega^2_2)}\big((\Gamma^2 + \Omega_1^2) (\Gamma\cosh{\Omega_2 t} + \Omega_2\sinh{\Omega_2 t}) \nonumber \\
 &- (\Gamma^2 - \Omega_2^2) (\Gamma\cos{\Omega_1 t} -\Omega_1\sin{\Omega_1 t} )\big).
\end{align}

In \cite{NBB1} it was shown that, for  the sharp redox potential, the ET efficiency is rather slow function of time. For instance, for $\varepsilon = 60 \rm \,ps^{-1}$ and $10 \rm\, ps < V < 40 \rm \,ps^{-1}$,  the ET efficiency approaches a value close to 1 for relatively large times, $t\gtrsim150$ ps. However, the situation is changes drastically for the flat redox potential. For example,  for $\varepsilon =0$, we obtain
\begin{align}
\begin{array}{l l l}
\Omega_2 =0, &\Omega_1^2 = V^2- \Gamma^2, & V>\Gamma,\\
\Omega_1 =0, &\Omega_2 = 0, & V=\Gamma \\
\Omega_1 =0, &\Omega_2^2 = \Gamma^2 - V^2, & \Gamma > V.
\end{array}
\end{align}

Using these relations,  we find  that at the exceptional point the ET efficiency behaves  as
\begin{align}
\eta(t) = 1- e^{-\Gamma t}.
\end{align}
For $V \ll \Gamma$ ( $V \gg \Gamma$) the asymptotic behavior of $\eta(t) $ is
\begin{align}
\label{Eq17e}
\eta(t) \sim\left \{ \begin{array}{ll}
 1-   e^{-\Gamma t}, &  V \gg \Gamma,  \\
  1- \displaystyle e^{- V^2t/2\Gamma} , & V \ll \Gamma.
  \end{array}
  \right.
\end{align}

\begin{figure}[tbh]
\scalebox{0.4}{\includegraphics{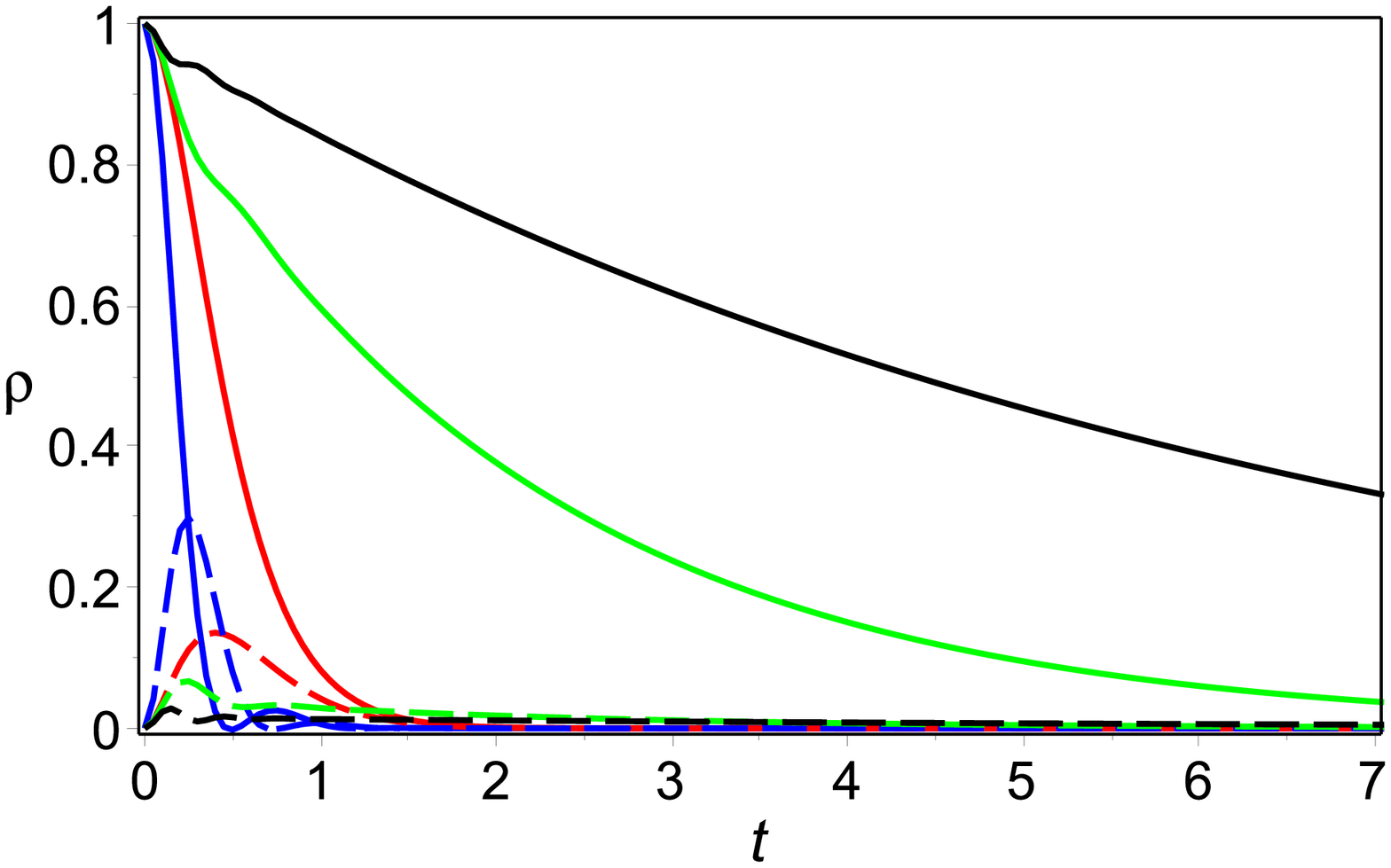}}
\scalebox{0.4}{\includegraphics{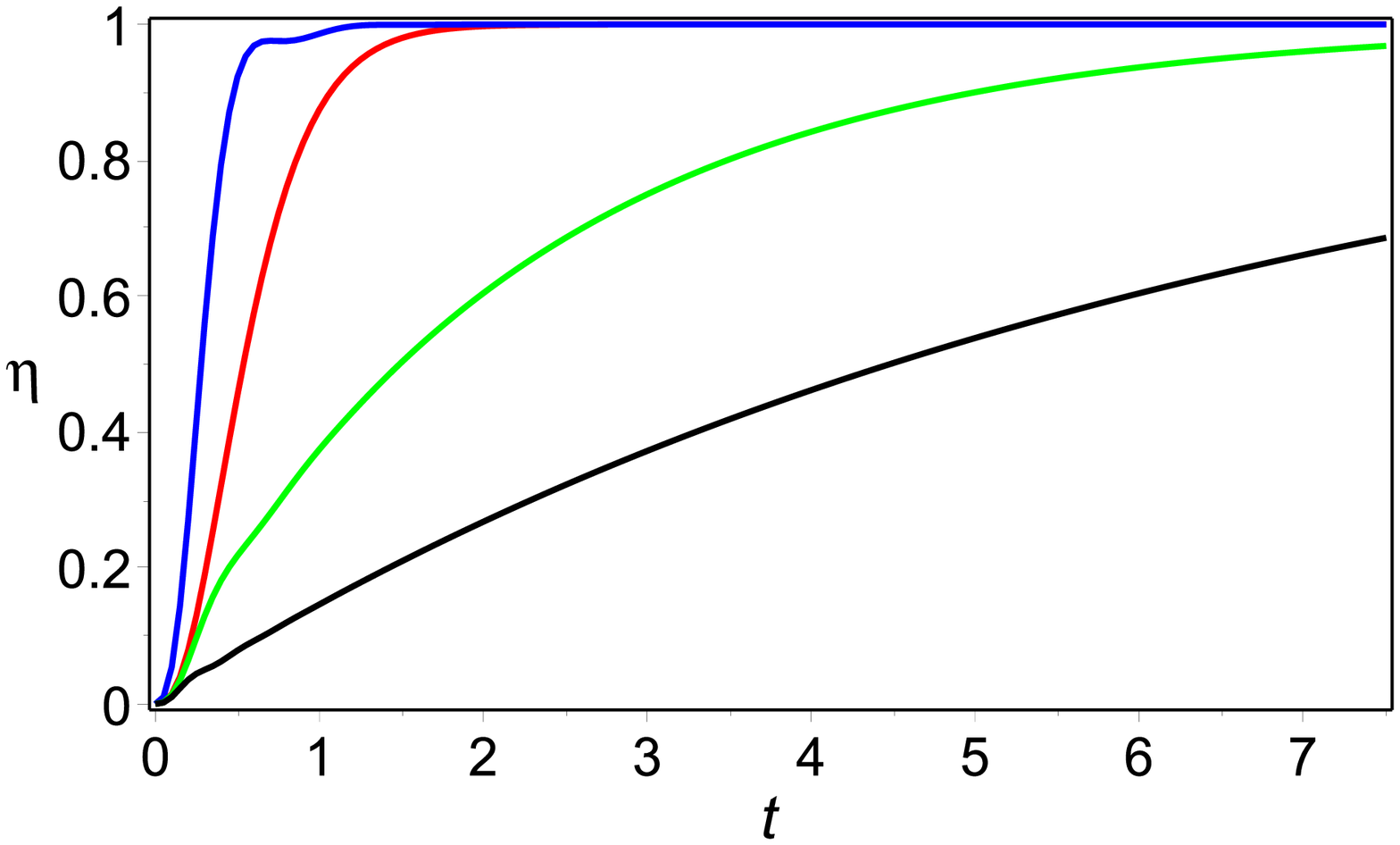}}
\caption{(Color online) Time dependence (in ps) of the sites population, $\rho_{11}(t)$ and $\rho_{22}(t)$, (top) and  ET efficiency (bottom) ($\Gamma= 5 \,\rm ps^{-1}  $). Solid and dashed lines correspond to, $\rho_{11}(t) $ and $\rho_{22}(t) $, respectively.  Blue line: $V= 10 \,\rm ps^{-1}, \, \varepsilon =0$;  red line (EP): $V= \Gamma=5 \,\rm ps^{-1}, \, \varepsilon =0$;  green line: $V= 10\, \rm ps^{-1}, \, \varepsilon =10\, \rm ps^{-1}$; black line: $V= 5 \,\rm ps^{-1}, \, \varepsilon =20 \,\rm ps^{-1}$.
\label{S2r}}
\end{figure}

 Comparing the obtained results with (\ref{Eq17r}),  we conclude  that the highest ET  rate is obtained for the flat redox potential  at $\varepsilon =0$, and $V \geq \Gamma$. This result can be interpreted by using a model of a single spin dynamics in an effective magnetic field. The above parameters are defined so that the effective magnetic field is oriented in the positive $x$-direction, and $V$ corresponds to the Rabi frequency--the frequency of rotation of the spin around the $x$-axis. Then, the above chosen conditions provide a rapid transition from the donor to the acceptor, with subsequent tunneling from the acceptor to the sink. The results of numerical simulations of the ET efficiency  are presented in Fig.~ \ref{S2r}.  One can see that, when $\varepsilon =0$ and $V\geq \Gamma$, the ET efficiency can approach a value close to 1 for short enough times, $\sim 2$ ps.

\section{Noise-assisted electron transfer to the sink}

In the presence of classical noise, the quantum dynamics of the ET can be described by the following effective non-Hermitian Hamiltonian (for details see Appendix A):
\begin{align} \label{Eq2b}
 \tilde{\mathcal H}=& \sum_n \varepsilon_n |n\rangle\langle  n |+ \sum_{m,n} \lambda_{mn}(t)|m\rangle\langle  n | \nonumber \\
& + \frac{V}{2}\sum_{m \neq n}  |m\rangle\langle  n | - i\frac{\Gamma_2}{2} |2\rangle \langle 2|, \quad m,n = 1,2,
\end{align}
where $\lambda_{mn}(t)$ describes the noise.  We denote  by $|1\rangle$ and $|2\rangle$ the donor and acceptor states in the site representation, respectively. The diagonal matrix elements of noise, $\lambda_{nn}$, are responsible for decoherence, and the off-diagonal matrix elements, $\lambda_{mn}$ ($ m \neq n$), lead to the relaxation processes. In what follows, we use a spin-fluctuator model of noise, modeling the noise by an ensemble of fluctuators \cite{GABS,BGA,NB1b}.

The evolution of the  average diagonal components of the density matrix is described by the following system of integro-differential equations (see Appendix B):
\begin{align} \label{IB4}
\frac{d}{dt}{\langle{\rho}}_{11}(t)\rangle =&- \int_0^t { K}(t,t')\big(\big\langle{\rho}_{11}(t')\big\rangle -\big\langle{\rho}_{22}(t')\big\rangle\big)dt', \\
\frac{d}{dt}{\langle{\rho}}_{22}(t)\rangle =&\int_0^t { K}(t,t')\big(\big\langle{\rho}_{11}(t')\big\rangle -\big\langle{\rho}_{22}(t')\big\rangle\big)dt' \nonumber \\
&- 2\Gamma \langle { \rho}_{22}(t)\rangle,
\label{IB5}
\end{align}
where the average $\langle\; \rangle$ is taken over the random process describing noise, and the kernel, $K(t,t')$, is given by
\begin{align}\label{K1}
  K(t,t') = e^{-\Gamma (t- t')}\big(\big\langle{\tilde V}_{21}(t){\tilde V}_{12}(t') \big\rangle + \big\langle{\tilde V}_{21}(t'){\tilde V}_{12}(t)\big\rangle\big).
\end{align}

In the rest of this paper, we restrict ourselves to considering only the diagonal noise effects, assuming that the noisy environment is the same for both the donor and the acceptor sites (collective noise). Then, one can write $\lambda_1(t) = g_1 \xi(t) $ and $\lambda_2(t) = g_2 \xi(t) $, where $g_{1,2}$ are the interaction constants, and $\xi(t)$ is a random variable describing a stationary noise with the correlation function,
$ \chi(t-t')=\langle \xi(t)\xi(t')\rangle,$
given by \cite{NB1b}
\begin{eqnarray}
\chi(\tau) = \sigma^2 A\Big(E_1(2\gamma_m \tau) - E_1(2\gamma_c \tau) \Big ), \quad \tau = |t-t'|.
\label{chi_4}
\end{eqnarray}
Here $E_n(z)$ denotes the Exponential integral \cite{abr}, $A= 1/\ln(\gamma_c/\gamma_m)$, $\sigma^2= \chi(0)$ and $\gamma_m$ and $\gamma_c$ ($\gamma_m\ll\gamma_c$) indicate the boundaries of the switching rates in the ensemble of random fluctuators. The correlation function includes, besides the amplitude, $\sigma$,  two fitting parameters: $\gamma_m$ and $\gamma_c$.  Taking into account available theoretical and experimental data \cite{TMT,CBC,JBS}, we have chosen for our numerical simulations the  parameters $\gamma_m$ and $\gamma_c$ as follows: $ 2\gamma_m = 10^{-4}\rm ps^{-1} $, $ 2\gamma_c = 1\rm ps^{-1} $.

Using the Gaussian approximation, we obtain the following expression for the kernel (see Appendix B):
\begin{align}
K(t -t') =\frac{V^2}{2} \cos(\varepsilon(t-t'))\exp\bigg(-\Gamma(t-t')- \frac{\langle\kappa^2(t-t')\rangle}{2}\bigg),
\label{Eq15b}
\end{align}
where
\begin{align}
\langle\kappa^2(t-t')\rangle=2 D^2\int^{t-t'}_0 d\tau'\int_{0}^{\tau'}d\tau''\chi(\tau' -\tau''),
\label{G2a}
\end{align}
and we denote $D =|g_1-g_2|$. The result of integration with the correlation function (\ref{chi_4})
is given by \cite{NB1b}
\begin{align}
 \langle \kappa^2(\tau) \rangle = &D^2  \sigma^2 A \bigg( \frac{E_{3}(2\gamma_m
 \tau)}{4\gamma_m^{2}}-  \frac{E_{3}(2\gamma_c\tau)}{4\gamma_c^{2}}  \nonumber \\
 & + \frac{\tau}{4\gamma_m^{2}} -\frac{\tau}{4\gamma_c^{2}} +\frac{1}{4\gamma_c^{2}}
  -\frac{1}{4\gamma_m^{2}}\bigg)
  \label{C7}
\end{align}

\begin{figure}[tbh]
\scalebox{0.4}{\includegraphics{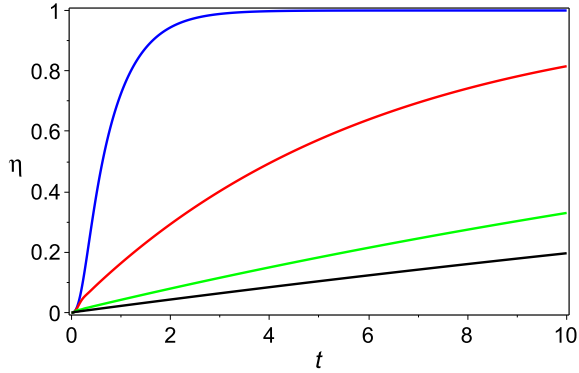}}
\caption{(Color online)
Time dependence (ps) of  the ET efficiency ($V=\Gamma = 1 \,\rm ps^{-1}$, $D\sigma = 5 \,\rm ps^{-1}$). Blue line: $\varepsilon= 0$ (EP), red line: $\varepsilon= 20 \,\rm ps^{-1}$, green line: $\varepsilon= 40 \,\rm ps^{-1}$, black line: $\varepsilon= 60 \,\rm ps^{-1}$.
\label{EP_5}}
\end{figure}

\begin{figure}[tbh]
\scalebox{0.4}{\includegraphics{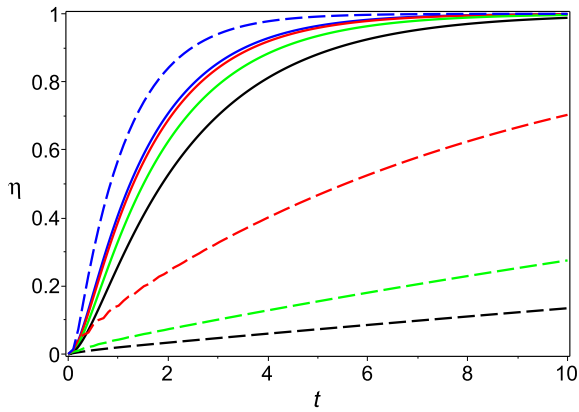}}
\caption{(Color online) Time dependence (in ps) of  the ET efficiency ($V=10 \,\rm ps^{-1}$, $\Gamma=1 \,\rm ps^{-1}$ ).  Solid lines correspond to $\eta(t)$ in the presence of noise with the amplitude $D\sigma = 40 \,\rm ps^{-1}$. Blue line: $\varepsilon= 0$, red line: $\varepsilon= 20 \,\rm ps^{-1}$, green line: $\varepsilon= 40 \,\rm ps^{-1}$, black line: $\varepsilon= 60 \,\rm ps^{-1}$. Dashed lines correspond to $\eta(t)$, in the absence of noise.
\label{NET_1}}
\end{figure}

In Figs.~\ref{EP_5} and \ref{NET_1} we present the results of numerical simulations for the tunneling rate, $\Gamma=1 \,\rm ps^{-1}$, and different parameters, $V$, $\varepsilon$, and the amplitude of noise, $D\sigma$. As one can see from Fig. ~\ref{EP_5}, a low level of the noise does not improve the ET efficiency rates. However, if the amplitude is sufficiently large, the noise significantly accelerates the ET to the sink (Fig. ~\ref{NET_1}).

\begin{figure}[tbh]
\scalebox{0.4}{\includegraphics{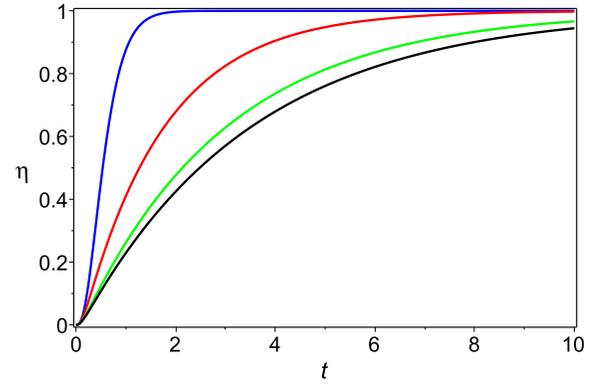}}
\caption{(Color online) Time dependence (in ps) of the  ET efficiency at the EP in the presence of noise ($ \varepsilon=0,  V = \Gamma=5 \,\rm ps^{-1} $). Blue line: $D\sigma= 0$, red line: $D\sigma= 20\, \rm ps^{-1}$, green line: $D\sigma= 40\, \rm ps^{-1}$, black line: $D\sigma= 60 \,\rm ps^{-1}$.
\label{S2a}}
\end{figure}

\begin{figure}[tbh]
\scalebox{0.375}{\includegraphics{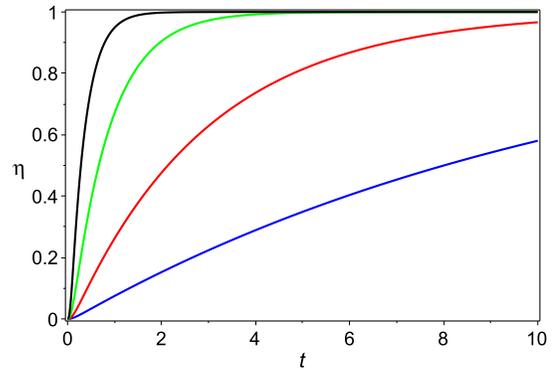}}
\caption{(Color online) Time dependence (in ps) of the ET efficiency in the vicinity of the EP in the presence of noise ($ \varepsilon=0,  D\sigma = 60 \,\rm ps^{-1},  \Gamma=5 \,\rm ps^{-1} $). Blue line: $V= 2.5\,\rm ps^{-1}$, red line: $V= 5 \,\rm ps^{-1}$ (EP), green line: $V= 10 \,\rm ps^{-1}$, black line: $V= 20 \,\rm ps^{-1}$.
\label{S2b}}
\end{figure}

In  \cite{NBB1} we show that for a sharp redox potential, noise can greatly improve the rate of ET to the sink.  Our numerical results presented in Fig. \ref{NET_1}, demonstrate that  this is true for any redox potential, if the noise is strong enough.
In Fig.~\ref{S2a}  we present the results of numerical simulations  at the EP and in presence of noise for the tunneling rate $\Gamma =5 \,\rm ps^{-1} $ ($ \varepsilon=0,  V = \Gamma=5 \,\rm ps^{-1} $).  As can be observed, at the exceptional point (EP) the noise decreases the rate of the ET.  The behavior of the system in the vicinity of the EP is rather complicated, and the ET efficiency is sensitive to the choice of parameters (Fig.~\ref{S2b} ). For instance, for the flat redox potential with $\varepsilon =0$ and an amplitude of noise, $D\sigma= 60 \,\rm ps^{-1}$, the ET efficiency  approaches a value close to 1 for short enough time, $\sim 2$ ps.

\subsection{Electron transport approximated by differential equations}

Under some conditions, the system of integro-differential equations,
\begin{align} \label{IB4a}
\frac{d}{dt}{\langle{\rho}}_{11}(t)\rangle =&- \int_0^t
 { K}(t,t')\big(\big\langle{\rho}_{11}(t')\big\rangle -\big\langle{\rho}_{22}(t')\big\rangle\big)dt', \\
\frac{d}{dt}{\langle{\rho}}_{22}(t)\rangle =&\int_0^t { K}(t,t')\big(\big\langle{\rho}_{11}(t')\big\rangle -\big\langle{\rho}_{22}(t')\big\rangle\big)dt' \nonumber \\
&- 2\Gamma \langle { \rho}_{22}(t)\rangle,
\label{IB5a}
\end{align}
can be approximated by the following system of the ordinary differential equations:
\begin{align} \label{BN6a}
\frac{d}{dt}{\langle{\rho}}_{11}(t)\rangle =&- {\mathfrak R}(t)\big(\big\langle{\rho}_{11}(t)\big\rangle -\big\langle{\rho}_{22}(t)\big\rangle\big) , \\
\frac{d}{dt}{\langle{\rho}}_{22}(t)\rangle = &\,{\mathfrak R}(t)\big(\big\langle{\rho}_{11}(t)\big\rangle -\big\langle{\rho}_{22}(t)\big\rangle\big)  - 2\Gamma \langle {\rho}_{22}(t)\rangle  ,
\label{BN7}
\end{align}
where $ {\mathfrak R}(t)= \int_{0}^{t} K(\tau) d\tau$.

Below  we obtain the conditions for which the exact system of integro-differential equations (\ref{IB4a}) - (\ref{IB5a}) can be approximated by Eqs. (\ref{BN6a}) - (\ref{BN7}). In the first order of the series expansion, we can write
\begin{align}
\langle{\rho}_{11}(t')\rangle \approx\langle{\rho}_{11}(t)\rangle -\frac{d}{dt}{\langle{\rho}}_{11}(t)\rangle (t-t') ,\\
\langle{\rho}_{22}(t')\rangle \approx\langle{\rho}_{22}(t)\rangle -\frac{d}{dt}{\langle{\rho}}_{22}(t)\rangle (t-t'),
\end{align}
where, in the same order, the derivatives on the r.h.s. are taken from Eqs. (\ref{BN6a}), (\ref{BN7}).

Using these results, after some transformations we find that  Eqs. (\ref{IB4a}) - (\ref{IB5a}) become
\begin{align} \label{B8}
\frac{d}{dt}{\langle{\rho}}_{11}(t)\rangle =&- {\mathfrak R}(t)\big(\big\langle{\rho}_{11}(t)\big\rangle -\big\langle{\rho}_{22}(t)\big\rangle\big) (1-{\mathfrak R}_1(t) ) \nonumber\\
& + 2\Gamma {\mathfrak R}_1(t) \langle {\rho}_{22}(t)\rangle,  \\
\frac{d}{dt}{\langle{\rho}}_{22}(t)\rangle = &\,{\mathfrak R}(t)\big(\big\langle{\rho}_{11}(t)\big\rangle -\big\langle{\rho}_{22}(t)\big\rangle\big) (1-{\mathfrak R}_1(t) ) \nonumber \\
&- 2\Gamma \langle {\rho}_{22}(t)\rangle (1-{\mathfrak R}_1(t)) ,
\label{B8a}
\end{align}
where ${\mathfrak R}_1(t)= \int_{0}^{t} \tau K(\tau) d\tau $. From here it follows that the system of integro-differential equations (\ref{IB4a}) -  (\ref{IB5a}) can be approximated by the system of the first order ordinary differential equations in the interval of time $0 < t < \infty$, if $|\int_{0}^{\infty} \tau K(\tau) d\tau  | \ll 1$.

Assuming that the correlation function, $\chi(t)$, is a rapidly decreasing function, as $t\rightarrow \infty$, we can approximate
\begin{align}
\exp\bigg(- D^2\int^{t}_0 d\tau\int_{0}^{\tau}ds\,\chi(\tau -s) \bigg)
 \approx
\exp\bigg(- \frac{(D\sigma t)^2}{2}\bigg).
\end{align}
\begin{figure}[tbh]
\scalebox{0.35}{\includegraphics{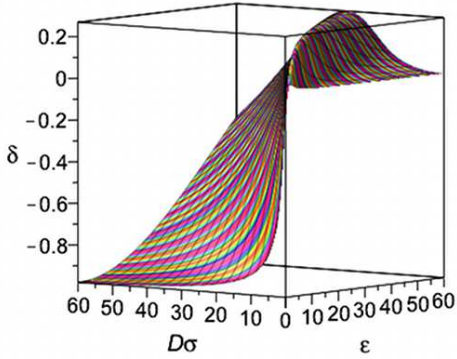}}
\caption{(Color online)  Dependence of  $\delta$ on $\varepsilon$ and $D\sigma$ ($\Gamma =  1\, \rm ps^{-1}$).
\label{VAP3d}}
\end{figure}

Performing the integration with the kernel
\begin{align}
K(\tau) =\frac{V^2}{2} \cos(\varepsilon \tau)\exp\bigg(-\Gamma \tau- \frac{(D\sigma)^2}{2}  \tau^2\bigg),
\end{align}
we obtain the following estimate:
\begin{align}
\bigg|\int_0^\infty K(\tau)\tau d\tau\bigg| =\bigg|\frac{V^2\delta}{2D^2\sigma^2}\bigg|< \frac{V^2 |\delta|_{\max}}{2D^2\sigma^2}
\ll 1,
\end{align}
where
\begin{align}
\delta=& -1+\frac{\sqrt{\pi}q}{4p}\exp\bigg(\frac{q^2}{4p^2}\bigg){\rm erfc}\bigg(\frac{q}{2p}\bigg)\nonumber \\
  &+\frac{\sqrt{\pi}\bar q}{4p} \exp\bigg(\frac{{\bar q}^2}{4p^2}\bigg){\rm erfc}\bigg(\frac{\bar q}{2p}\bigg),
\label{BR}
\end{align}
where $p=D\sigma/\sqrt{2}$, $q=\Gamma+ i\varepsilon$, $\bar q=\Gamma-  i\varepsilon$, and ${\rm erf}c(z)$ denotes the complementary error function\cite{abr}.

 Using the properties of the ${\rm erfc} $ function, one can show that $|\delta|\leq 1$  for any choice of parameters $\Gamma, \varepsilon$ and the amplitude of noise, $D\sigma$ (Fig. \ref{VAP3d}).   Consequently, the  condition of validaty of the approximation (\ref{BN6a}) - (\ref{BN7}) can be written as $V \ll D\sigma$. This rough estimate can be improved greatly for the high level of noise leading to $V\leq D\sigma$.

\begin{figure}[tbh]
\scalebox{0.325}{\includegraphics{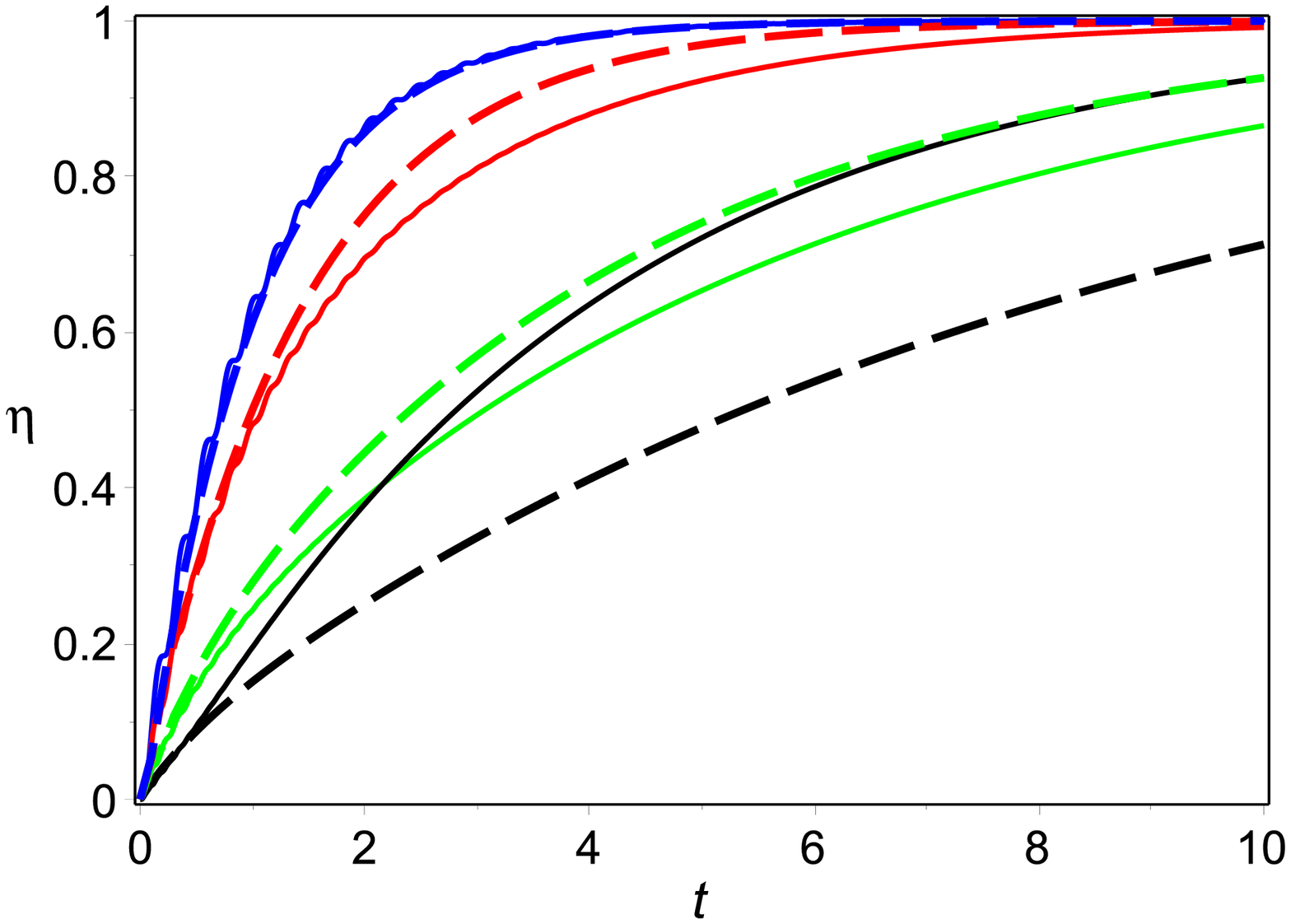}}
\scalebox{0.325}{\includegraphics{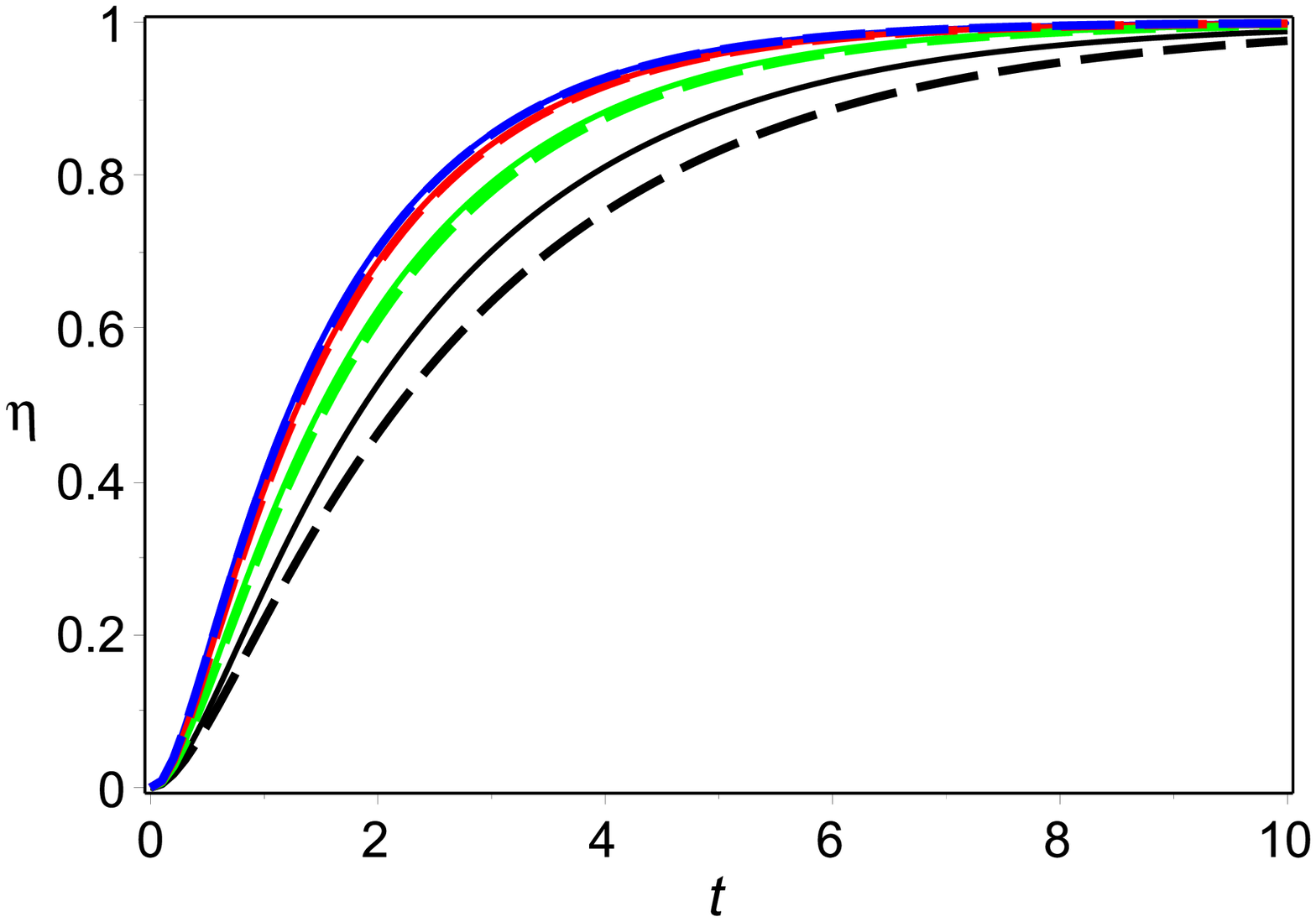}}
\caption{(Color online)  Time dependence (in ps) of  the ET efficiency, $\eta(t)$.  The solutions of the system of integro-differential equations are presented by solid lines ($\Gamma = 1 \,\rm ps^{-1}$). Dashed lines correspond to the solutions of the approximate system of differential equations. Blue line: $\varepsilon= 0$, red line: $\varepsilon= 20 \,\rm ps^{-1}$, green line: $\varepsilon= 40 \,\rm ps^{-1}$, black line: $\varepsilon= 60 \,\rm ps^{-1}$. Top: $V=30 \,\rm ps^{-1}$, $D\sigma = 5 \,\rm ps^{-1}$. Bottom: $V=10 \,\rm ps^{-1}$, $D\sigma = 40 \,\rm ps^{-1}$.
\label{NET_2}}
\end{figure}

\begin{figure}[tbh]
\scalebox{0.3}{\includegraphics{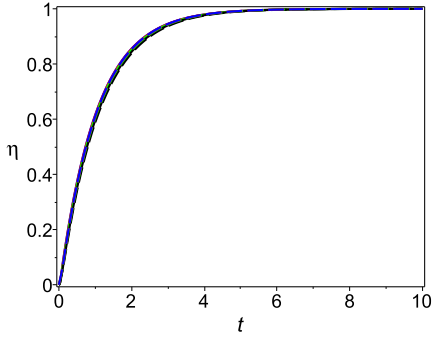}}
\caption{(Color online) Time dependence (in ps) of  the ET efficiency. Solid lines correspond to $\eta(t)$ obtained from the system of integro-differential Eqs. (\ref{IB4})-(\ref{IB5}) and dashed lines present the results for $\eta(t)$ obtained from the system of differential  Eqs. (\ref{BN6a})-(\ref{BN7}). Blue line: $\varepsilon= 0$, red line: $\varepsilon= 20 \,\rm ps^{-1}$, green line:
$\varepsilon= 40 \,\rm ps^{-1}$, black line: $\varepsilon= 60 \,\rm ps^{-1}$ ($\Gamma= 1\, \rm ps^{-1}$, $V= D\sigma=40 \,\rm ps^{-1}$).  One can observe the excellent agreement between the solutions.
\label{IDF_1}}
\end{figure}

In Figs. \ref{NET_2}, \ref{IDF_1} we present the results of our numerical simulations for $\Gamma= 1\, \rm ps^{-1}$.  As one can see, for  $V\leq D\sigma$ there is good agreement between the solutions obtained from the system of integro-differential Eqs. (\ref{IB4})-(\ref{IB5}) and the approximate system of differential  Eqs. (\ref{BN6a})-(\ref{BN7}).

The advantage of using differential equations instead of integro-differential equations is (i) the ability to introduce effective rates and (ii)  the ability to compare the results with the Markus theory. The computation of the asymptotic ET, $ {\mathfrak R}_\Gamma  = \lim_{t\rightarrow\infty} {\mathfrak R}(t)$,  yields \cite{NBB1}
\begin{align}
 {\mathfrak R}_{\Gamma} =&\frac{V^2\sqrt{2\pi}}{8D\sigma} \Bigg( \exp\bigg(\frac{(\Gamma+ i\varepsilon)^2}{2D^2 \sigma^2}\bigg){\rm erfc}\bigg(\frac{\Gamma+ i\varepsilon}{\sqrt{2}D \sigma}\bigg)\nonumber \\
  &+ \exp\bigg(\frac{(\Gamma - i\varepsilon)^2}{2D^2 \sigma^2}\bigg){\rm erfc}\bigg(\frac{\Gamma - i\varepsilon}{\sqrt{2}D \sigma}\bigg)\Bigg ).
\label{Eq20a}
\end{align}

\begin{figure}[tbh]
\scalebox{0.375}{\includegraphics{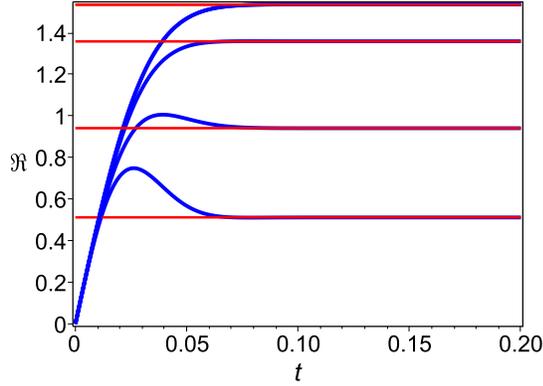}}
\caption{(Color online) Blue line describes the time dependence of the ET rate,  $ {\mathfrak R}(t)$. Red line corresponds to the asymptotic rate, ${\mathfrak R}_{\Gamma}$, given by Eq. (\ref{Eq20a}).  From top to bottom: $\varepsilon = 0,\,20,\,40,\,60\, \rm ps^{-1}$ ($\Gamma =1\, \rm ps^{-1}$, $V = 10\, \rm ps^{-1}$ and $D\sigma= 40 \,\rm ps^{-1}$).
\label{RG}}
\end{figure}
\begin{figure}[tbh]
\scalebox{0.35}{\includegraphics{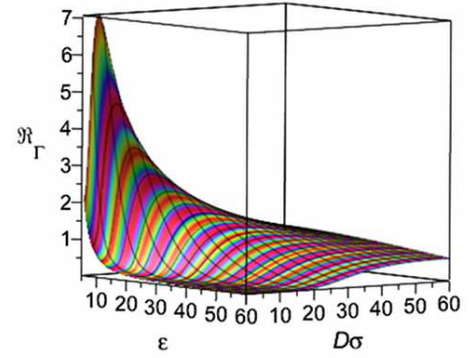}}
\caption{(Color online)  Dependence of the asymptotic rate, ${\mathfrak R}_{\Gamma}$, on $\varepsilon$ and $D\sigma$ ($V =  10\, \rm ps^{-1}$, $\Gamma =  1\, \rm ps^{-1}$).
\label{RG3d}}
\end{figure}
In Fig. \ref{RG}, we compare the results of numerical calculations of the relaxation rate, ${\mathfrak R}(t)$ (blue line), with the asymptotic expression ${\mathfrak R}_{\Gamma}$ (red line) for different choices of parameters. As  can see in Fig. \ref{RG3d}, for the given $V$  and $\varepsilon$, the rate ${\mathfrak R}_{\Gamma}$ reaches its maximum value when the amplitude of noise $D\sigma \approx \varepsilon$.

Inserting ${\mathfrak R}_\Gamma $ instead of ${\mathfrak R}(t)$ into  Eqs. (\ref{BN6a}) and (\ref{BN7}), we obtain the following system of the  differential equations:
\begin{align} \label{TN6b}
\frac{d}{dt}{\langle{\rho}}_{11}(t)\rangle =&-{\mathfrak R}_{\Gamma}\big(\big\langle{\rho}_{11}(t)\big\rangle -\big\langle{\rho}_{22}(t)\big\rangle\big) , \\
\frac{d}{dt}{\langle{\rho}}_{22}(t)\rangle = &\,{\mathfrak R}_{\Gamma}\big(\big\langle{\rho}_{11}(t)\big\rangle -\big\langle{\rho}_{22}(t)\big\rangle\big)  - 2\Gamma \langle {\rho}_{22}(t)\rangle .
\label{TN7b}
\end{align}
The solution is given by
\begin{align}\label{AS1r}
{\langle{\rho}}_{11}(t)\rangle = &\bigg(\frac{1}{2} - \frac{\Gamma}{2\sqrt{ {\mathfrak R}_\Gamma^2 + \Gamma^2}}\bigg)e^{-{\mathfrak R}_1t}\nonumber \\
&+  \bigg(\frac{1}{2} + \frac{\Gamma}{2\sqrt{ {\mathfrak R}_\Gamma^2 + \Gamma^2}}\bigg)e^{-{\mathfrak R}_2t}, \\
{\langle{\rho}}_{22}(t)\rangle =& \frac{ {\mathfrak R}_\Gamma}{2\sqrt{ {\mathfrak R}_\Gamma^2 + \Gamma^2}}\bigg(e^{-{\mathfrak R}_2t} -e^{-{\mathfrak R}_1t}  \bigg),
\label{AS2}
\end{align}
where ${\mathfrak R}_{1,2} =  {\mathfrak R}_\Gamma + \Gamma \pm \sqrt{ {\mathfrak R}_\Gamma^2 + \Gamma^2} $. The computation of the ET efficiency yields \cite{NBB1}
\begin{widetext}
\begin{align}\label{Eq21}
\eta(t) = 1- e^{-\frac{({\mathfrak R}_1 + {\mathfrak R}_2)t}{2}}\bigg( \cosh\frac{({\mathfrak R}_1- {\mathfrak R}_2)t}{2}  + \frac{{\mathfrak R}_1 + {\mathfrak R}_2}{{\mathfrak R}_1 - {\mathfrak R}_2} \sinh\frac{({\mathfrak R}_1- {\mathfrak R}_2)t}{2} \bigg).
\end{align}
\end{widetext}
Its asymptotic behavior is
\begin{align}\label{Eq21a}
\eta(t) \approx 1- \frac{{\mathfrak R}_1 }{{\mathfrak R}_1 - {\mathfrak R}_2} e^{- {\mathfrak R}_2 t}.
\end{align}
As one can see from Eq. (\ref{Eq21}), there are two ET rates, ${\mathfrak R}_1$ and ${\mathfrak R}_2$. However, the asymptotic behavior of the ET efficiency is defined by the lowest ET rate, ${\mathfrak R}_2$.

\begin{figure}[tbh]
\scalebox{0.2}{\includegraphics{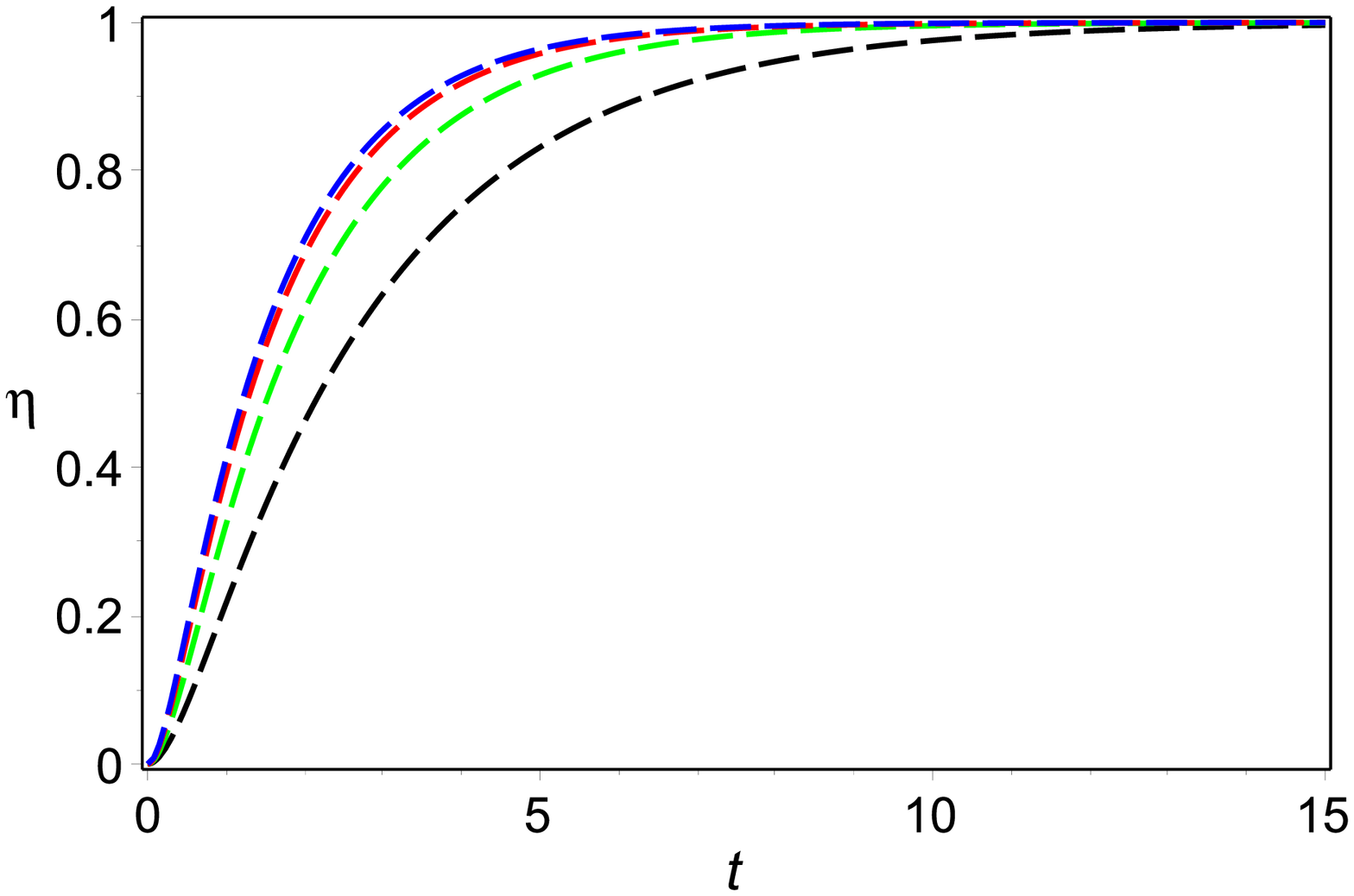}}
\scalebox{0.205}{\includegraphics{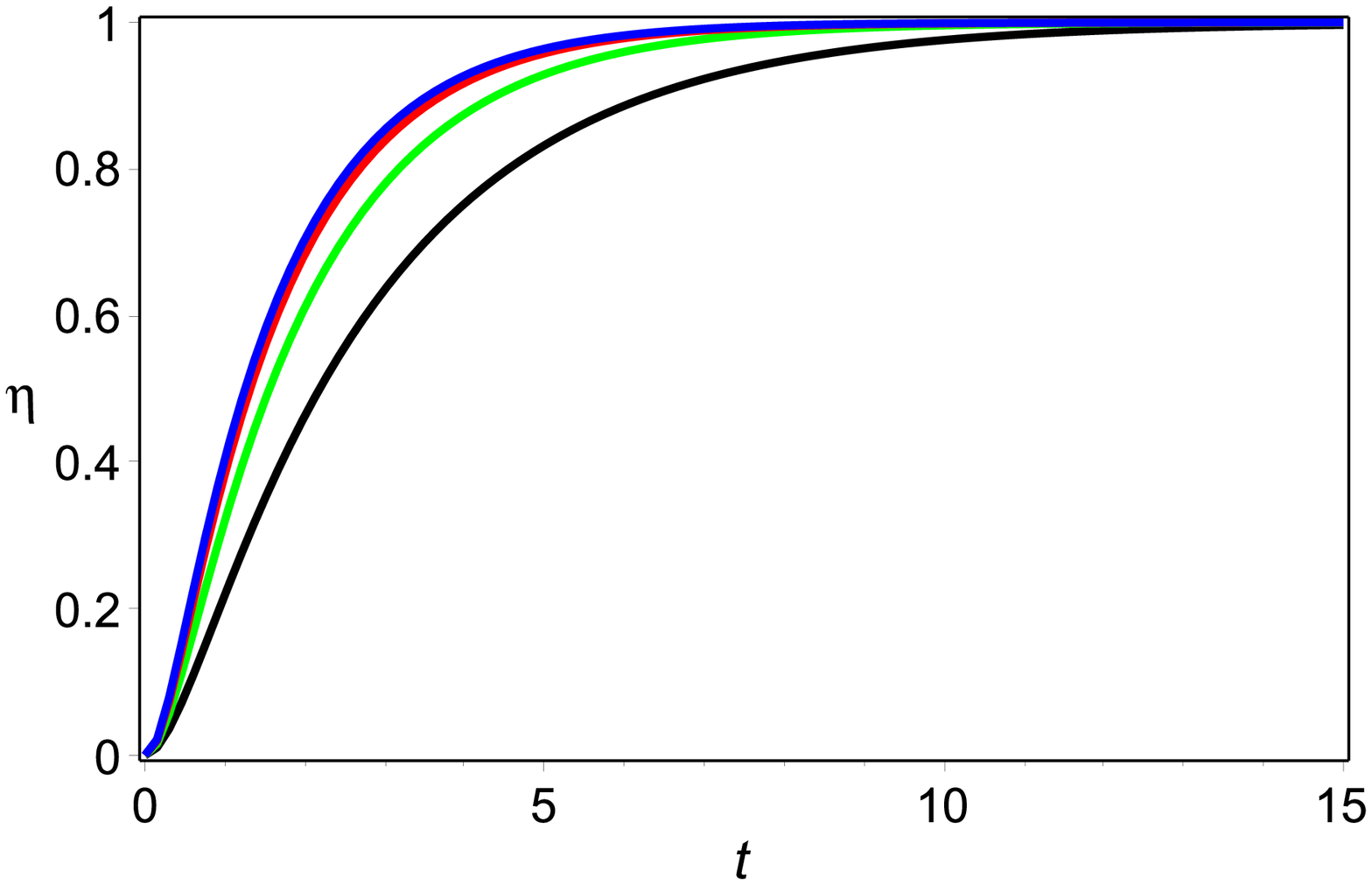}}
\scalebox{0.325}{\includegraphics{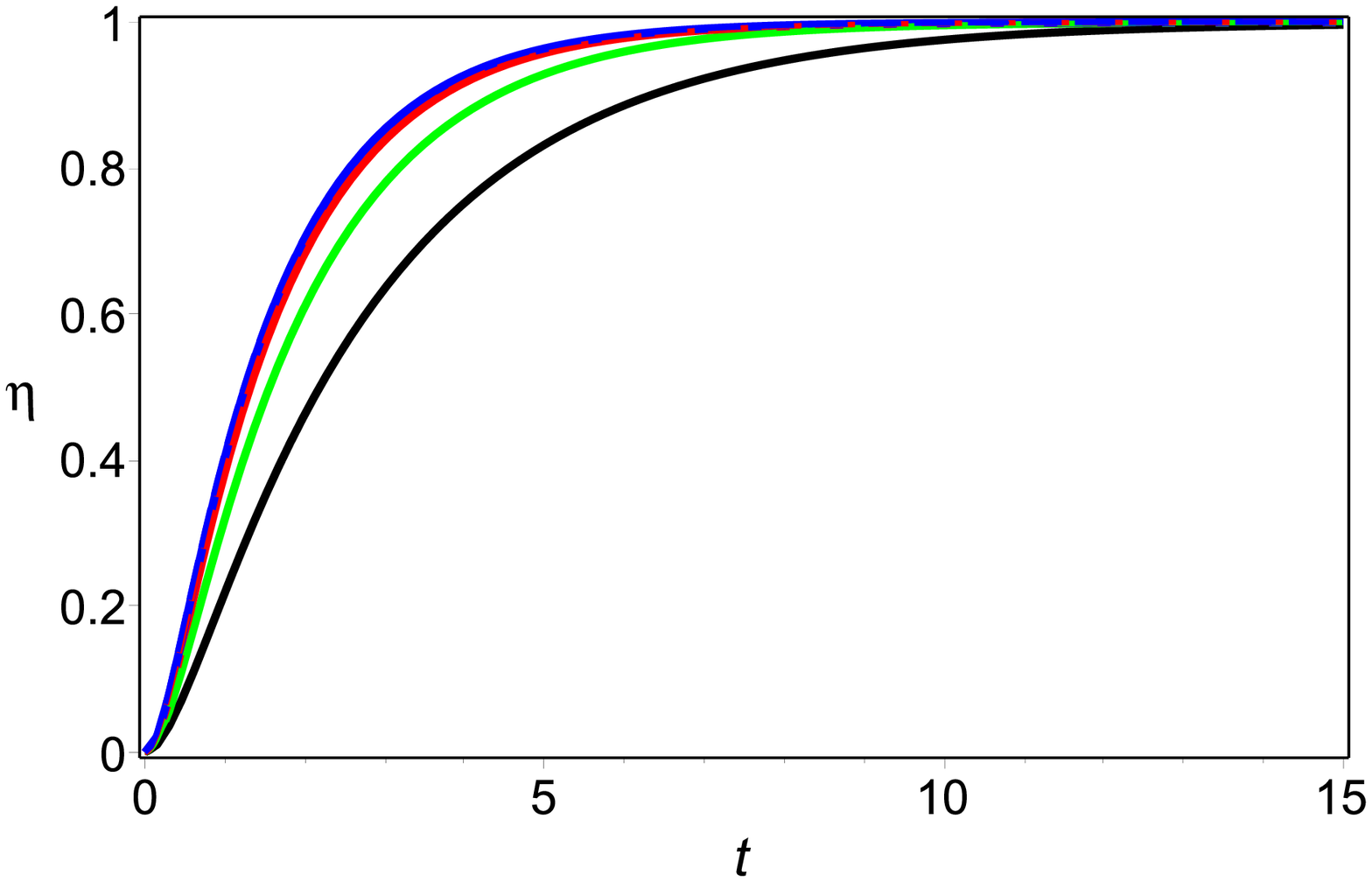}}
\caption{(Color online)
Time dependence (in ps) of  the ET efficiency, $\eta(t)$ ($\Gamma = 1 \,\rm ps^{-1}$, $V=10 \,\rm ps^{-1}$, $D\sigma = 40 \,\rm ps^{-1}$). Top. Left: analytical solution (dashed lines). Right: results of numerical simulations (solid lines). Blue line: $\varepsilon= 0$, red line: $\varepsilon= 20 \,\rm ps^{-1}$, green line: $\varepsilon= 40 \,\rm ps^{-1}$, black line: $\varepsilon= 60 \,\rm ps^{-1}$.
Bottom. The both graphics are overlapping.
\label{DF_2}}
\end{figure}

As shown in Fig. \ref{RG}, the ET rate reaches its asymptotic value, ${\mathfrak R}(t) \rightarrow {\mathfrak R}_{\Gamma}$, quite rapidly, at $t\approx 0.1 \, \rm ps$. This allows us to use the analytical solutions to describe tunneling to the sink with very high degree of accuracy. This conclusion is confirmed by our numerical simulations presented in Fig. \ref{DF_2}. One can observe the excellent agreement between the ET efficiency given by formula (\ref{Eq21}) and the results obtained from Eqs. (\ref{BN6a}) - (\ref{BN7}).

\section{Modified model with two sinks}

In this Section, we generalize our model by including two sinks interacting independently with the donor and acceptor. (See Fig. 13.) The main reasons for this generalization are the following.  When using a single sink which interacts with the acceptor, the asymptotic for large times of the ET efficiency, $\eta(t)$,   approaches the unit independently of the parameters of the system, and  the tunneling rate, $\Gamma$.  The additional sink, which interacts with the donor, describes the leakage of the electron from the RC in the process of the ET. By manipulating the densities of the donor and acceptor states, or the tunneling rates, $\Gamma_1$ and $\Gamma_2$, (see below), one can fit the asymptotic behavior of the ET efficiency, $\eta_2$, on the acceptor $(0\leq \eta_2\leq 1)$ with its experimental value. This approach is used for modeling the dynamics of the ET in photosynthetic complexes. (See, for example,  \cite{PPM,PPM1}, and references therein.)

The Hamiltonian of the system can be written as
\begin{align}
H_t = E_d|d\rangle \langle d|+
E_a|a\rangle \langle a| + \frac{V}{2}(|d\rangle \langle a|+ |a\rangle \langle d|) \nonumber \\
+\sum^{N_d}_{i=1}\big (E_i |i\rangle\langle  i | +  V_{di}|d\rangle\langle  i | + V_{id}|i\rangle\langle d |\big) \nonumber \\
+\sum^{N_a}_{j=1}\big (E_j |j\rangle\langle  j| +  V_{aj}|a\rangle\langle  j | + V_{ja}|j\rangle\langle a |\big),
\label{TH}
\end{align}
where $E_i(E_j)$ are energy levels of the sinks coupled with the donor (acceptor), and $V_{da} =V/2$.

After the transition to the continuum spectra of the sinks, the system is governed by the effective non-Hermitian Hamiltoniancan, $ \tilde{\mathcal H}= {\mathcal H}- i \mathcal W$, where
 \begin{align}
 {\mathcal H} =& \varepsilon_1|1\rangle \langle 1|+ \varepsilon_2|2\rangle \langle 2|+ \frac{V}{2}(|1\rangle \langle 2|+ |2\rangle \langle 1|),
 \label{Th1}
 \end{align}
where ${\mathcal H}$ is the dressed donor-acceptor Hamiltonian and
  \begin{align}
 \mathcal W = &\frac{1}{2}(\Gamma_1|1\rangle \langle 1|+\Gamma_2|2\rangle \langle 2|).
  \label{Th2}
 \end{align}
Passing from Eq. (\ref{TH})  to  Eqs. (\ref{Th1}) - (\ref{Th2}) we have changed $|d\rangle \rightarrow|1\rangle $ and $|a\rangle \rightarrow|2\rangle $ .  (See Appendix A.)

\begin{figure}[tbh]
\scalebox{0.3}{\includegraphics{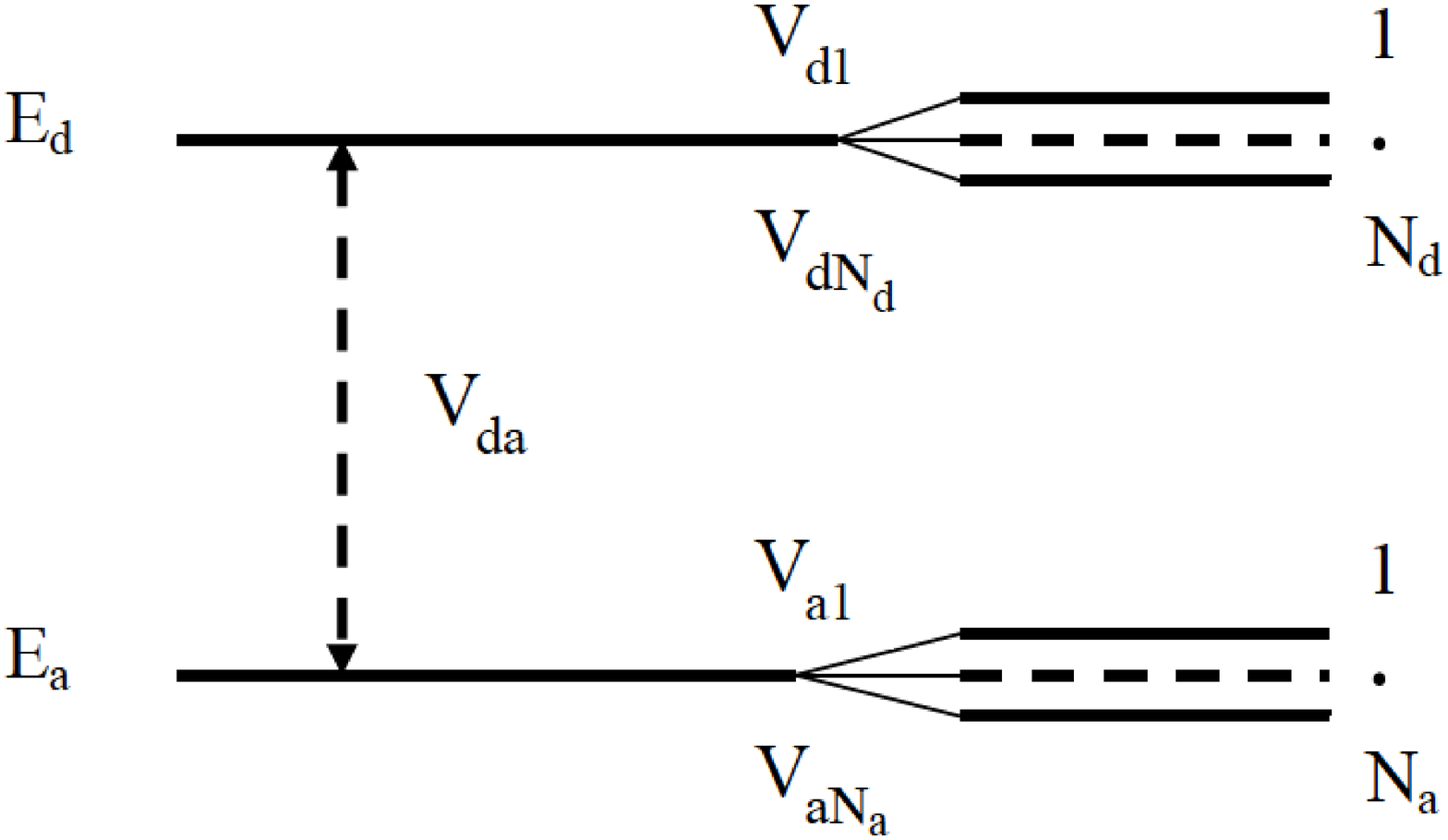}}
\caption{Schematic of our modified model consisting of donor and acceptor discrete energy levels, with the donor and acceptor coupled to independent sink reservoirs with nearly continuous spectrum.
\label{S2}}
\end{figure}

The  dynamics of the system is described by the Liouville  equation,
\begin{eqnarray}\label{L1}
   i \dot{ \rho} = [\mathcal H,\rho] - i\{\mathcal W,\rho\},
\end{eqnarray}
Further, we assume that initially the electron occupies the upper level (donor),  $\rho_{11}(0)=1$ and $\rho_{22}(0)=0$.  With these initial conditions, the solution of the Eq. (\ref{L1}) for the diagonal component of the density matrix is given by
\begin{align}
\rho_{11}(t)  =& {e^{-\Gamma t}} \bigg |\Big(\cos\frac{\Omega t}{2} - i\cos\theta\sin\frac{\Omega t}{2}\Big)\bigg|^2, \\
\rho_{22}(t)  = &{e^{-\Gamma t}} \bigg |\sin\theta\sin\frac{\Omega t}{2}\bigg |^2,
\label{P3dr}
\end{align}
where $\Gamma= (\Gamma_1 + \Gamma_2)/2$, $\Omega= \sqrt{V^2 +(\varepsilon + i\Gamma )^2}$  being the complex Rabi frequency, $\cos\theta = (\varepsilon + i\Gamma )/\Omega$, and $\sin\theta = V/\Omega$.

Setting $\Omega= \Omega_1 +i \Omega_2 $,  we obtain for $\rho_{22}(t) $ the simple analytical expression:
\begin{align}
\rho_{22}(t)  = \frac{  V^2 e^{-\Gamma t}}{2(\Omega^2_1 +\Omega^2_2)} \big (\cosh{\Omega_2 t} - \cos{\Omega_1 t} \big ).
\label{P3e}
\end{align}

We define the ET efficiency of trapping the electron in the acceptor's sink  as
\begin{eqnarray}
\eta_{2}(t) = \Gamma_2 \int_0^t \rho_{22}(\tau)d \tau.
\label{Eq16a}
\end{eqnarray}
Inserting $\rho_{22}(t)$ into (\ref{Eq16a}) and performing the integration, we obtain
\begin{widetext}
\begin{align}\label{Eq17}
 \eta_2(t) =\frac{\Gamma_2}{\Gamma_1 + \Gamma_2}\bigg( 1- \frac{ e^{-\Gamma t}}{\Gamma(\Omega^2_1 +\Omega^2_2)}\big((\Gamma^2 + \Omega_1^2) (\Gamma\cosh{\Omega_2 t} + \Omega_2\sinh{\Omega_2 t})
 - (\Gamma^2 - \Omega_2^2) (\Gamma\cos{\Omega_1 t} -\Omega_1\sin{\Omega_1 t} )\big)\bigg).
\end{align}
\end{widetext}
From here it follows  $\eta_2(t) \rightarrow \eta_{0}$, as $t \rightarrow \infty$, where
\begin{align}
\eta_{0} = \frac{\Gamma_2}{\Gamma_1 + \Gamma_2}.
\label{ET0}
\end{align}

\subsection{Noise-assisted electron transfer }

In the presence of classical diagonal noise, described by $\lambda_n =g_n \xi(t)$, the effective non-Hermitian Hamiltonian can be written as
\begin{align} \label{Eq2br}
 \tilde{\mathcal H}=& \sum_n \bigg(\varepsilon_n + g_n \xi(t)- i\frac{\Gamma_n}{2} \bigg) |n\rangle\langle  n | + \frac{V}{2}\sum_{m \neq n}  |m\rangle\langle  n |.
\end{align}

We describe the evolution of the system by the following  differential equations:
\begin{align} \label{RS1}
\frac{d}{dt}{\langle{\rho}}_{11}\rangle =&-{\mathfrak R}_{\Gamma}\big(\big\langle{\rho}_{11}\big\rangle -\big\langle{\rho}_{22}\big\rangle\big) -\Gamma_1 \langle {\rho}_{11}\rangle, \\
\frac{d}{dt}{\langle{\rho}}_{22}\rangle = &\,{\mathfrak R}_{\Gamma}\big(\big\langle{\rho}_{11}\big\rangle -\big\langle{\rho}_{22}\big\rangle\big)  - \Gamma_2 \langle {\rho}_{22}\rangle ,
\label{RS2}
\end{align}
where ${\mathfrak R}_\Gamma  = \lim_{t\rightarrow\infty} {\mathfrak R}(t)$ is given by
\begin{align}
 {\mathfrak R}_{\Gamma} =&\frac{V^2\sqrt{2\pi}}{8D\sigma} \Bigg( \exp\bigg(\frac{(\Gamma+ i\varepsilon)^2}{2D^2 \sigma^2}\bigg){\rm erfc}\bigg(\frac{\Gamma+ i\varepsilon}{\sqrt{2}D \sigma}\bigg)\nonumber \\
  &+ \exp\bigg(\frac{(\Gamma - i\varepsilon)^2}{2D^2 \sigma^2}\bigg){\rm erfc}\bigg(\frac{\Gamma - i\varepsilon}{\sqrt{2}D \sigma}\bigg)\Bigg ).
\label{Eq20ar}
\end{align}

The motivation to use this simplified description is as follows. As was shown in Sec. III,  approximation of integro-differential equations by  the system of differential equations is valid for $v\leq D\sigma$. In addition, since the ET rate reaches its asymptotic value, ${\mathfrak R}(t) \rightarrow {\mathfrak R}_{\Gamma}$, quite rapidly, at $t\approx 0.1 \, \rm ps$, we can use the asymptotic rates, ${\mathfrak R}_{\Gamma}$, instead of ${\mathfrak R}(t)$. This allows us to use the analytical solutions to describe tunneling to the sinks with very high degree of accuracy.

The solution of Eqs. (\ref{RS1}) - (\ref{RS2}), with the initial conditions, $\langle{\rho}_{11}(0)\big\rangle =1$ and $\langle{\rho}_{22}(0)\big\rangle =0$, is
\begin{align}\label{AS1r2}
{\langle{\rho}}_{11}(t)\rangle = &\bigg(\frac{1}{2} - \frac{\Delta}{2\sqrt{ {\mathfrak R}_\Gamma^2 + \Delta^2}}\bigg)e^{-{\mathfrak R}_1t}\nonumber \\
&+  \bigg(\frac{1}{2} + \frac{\Delta}{2\sqrt{ {\mathfrak R}_\Gamma^2 + \Delta^2}}\bigg)e^{-{\mathfrak R}_2t}, \\
{\langle{\rho}}_{22}(t)\rangle =& \frac{ {\mathfrak R}_\Gamma}{2\sqrt{ {\mathfrak R}_\Gamma^2 + \Delta^2}}\bigg(e^{-{\mathfrak R}_2t} -e^{-{\mathfrak R}_1t}  \bigg),
\label{AS2r}
\end{align}
where ${\mathfrak R}_{1,2} =  {\mathfrak R}_\Gamma + \Gamma \pm \sqrt{ {\mathfrak R}_\Gamma^2 + \Delta^2} $ and $\Delta =(\Gamma_2 -\Gamma_1)/2$.

The computation of the ET efficiency of tunneling  in the acceptor's sink yields
\begin{widetext}
\begin{align}\label{Eq21r}
\eta_2(t) =\frac{\Gamma_2{\mathfrak R}_\Gamma }{{\mathfrak R}_1 {\mathfrak R}_2}\Bigg (1- e^{-\frac{({\mathfrak R}_1 + {\mathfrak R}_2)t}{2}}\bigg( \cosh\frac{({\mathfrak R}_1- {\mathfrak R}_2)t}{2}  + \frac{{\mathfrak R}_1 + {\mathfrak R}_2}{{\mathfrak R}_1 - {\mathfrak R}_2} \sinh\frac{({\mathfrak R}_1- {\mathfrak R}_2)t}{2} \bigg)\Bigg).
\end{align}
\end{widetext}
From here it follows, as $t \rightarrow \infty$,
\begin{align}
\eta_2(t) \rightarrow \eta_{r} = \frac{\Gamma_2{\mathfrak R}_\Gamma }{{\mathfrak R}_1 {\mathfrak R}_2}.
\end{align}
Comparing the obtained results with the ET efficiency without noise, $\eta_0=\Gamma_2/(\Gamma_1 + \Gamma_2)$, (see Eq. (\ref{ET0})), we obtain
\begin{align}
 \eta_{r} =  \eta_{0}\frac{2\Gamma{\mathfrak R}_\Gamma }{2\Gamma{\mathfrak R}_\Gamma +\Gamma^2 - \Delta^2}.
 \label{RC1}
\end{align}
This can be represented in the form
\begin{align}
 \eta_{r} =  \eta_{0}\frac{(\Gamma_1+ \Gamma_2){\mathfrak R}_\Gamma }{(\Gamma_1+ \Gamma_2){\mathfrak R}_\Gamma +\Gamma_1\Gamma_2}.
\end{align}
From here it follows, that when ${\mathfrak R}_\Gamma  \gg \Gamma_1\Gamma_2/(\Gamma_1+\Gamma_2) $ the ET efficiency  $\eta_{r} \approx  \eta_{0}$. Generally, the ET efficiency with the presence of noise cannot exceed the ET efficiency without noise, $\eta_{r}\leq \eta_0$.
\begin{figure}[tbh]
\scalebox{0.215}{\includegraphics{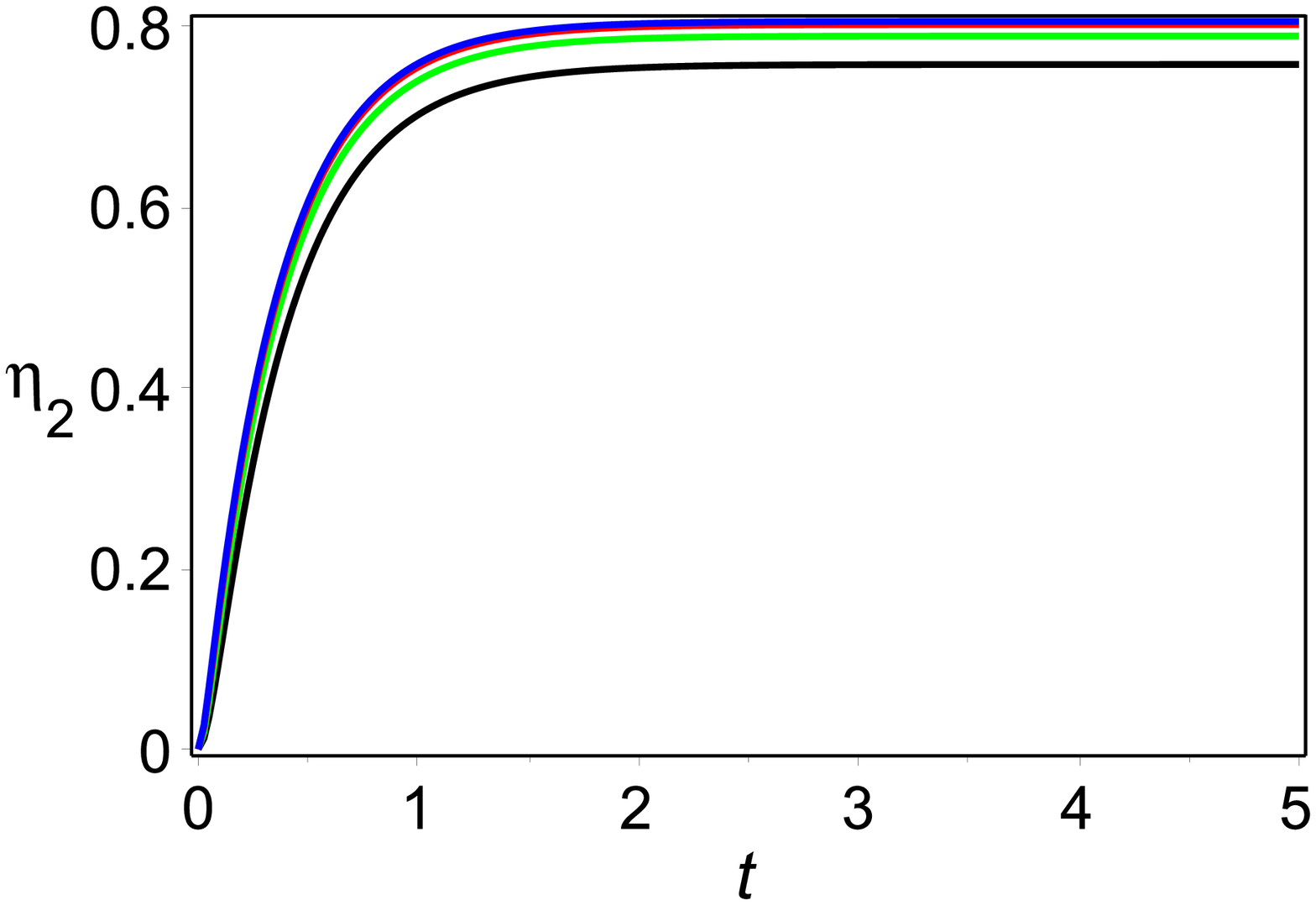}}
\scalebox{0.21}{\includegraphics{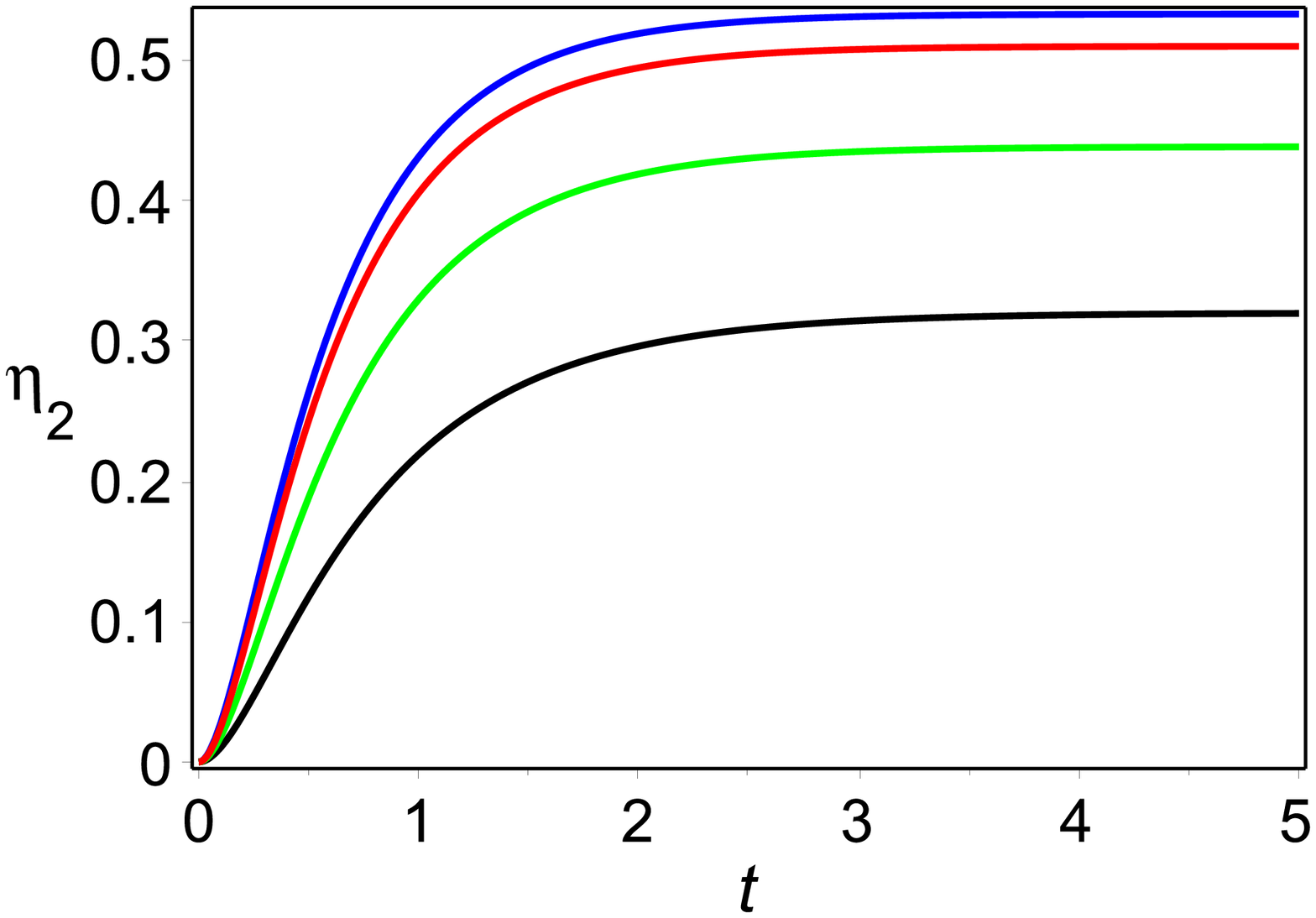}}
\caption{(Color online)
Time dependence (in ps) of  the ET efficiency, $\eta_2(t)$ ($\Gamma_1 = 1 \,\rm ps^{-1}$, $\Gamma_2 = 5 \,\rm ps^{-1}$).  Blue line: $\varepsilon= 0$, red line: $\varepsilon= 20 \,\rm ps^{-1}$, green line: $\varepsilon= 40 \,\rm ps^{-1}$, black line: $\varepsilon= 60 \,\rm ps^{-1}$. Left: $V=40 \,\rm ps^{-1}$, $D\sigma = 40 \,\rm ps^{-1}$. Right: $V=10 \,\rm ps^{-1}$, $D\sigma = 40 \,\rm ps^{-1}$.
\label{ETA}}
\end{figure}

In Fig. \ref{ETA}, we present the results of numerical simulations of the ET efficiency when two sinks are taken into account. As one can see from Eq. (\ref{RC1}), the asymptotic value for the ET efficiency, for chosen values of $\Gamma_{1,2}$, must satisfy the condition, $\eta_{rc}\leq\eta_0\approx 83.3\%$. As one can see, the  asymptotic value of $\eta_{rc}$ is close enough to $\eta_0$, for chosen in Fig. \ref{ETA} (left), parameters corresponding to black line.  We also would like to mention that, as the results presented in Fig. \ref{ETA} (left) demonstrate,  the saturation of $\eta_2(t)$ happens fast enough, at $t\approx 2\, \rm ps$.

\section{Conclusion and discussions}

In this paper, we  analyzed analytically and numerically the simplest model of the electron transfer  (ET) between two protein sites, donor and acceptor, in the presence of  classical (external) noise that is characterized by its amplitude and its correlation time. The noise is described by the well-known model of two-level fluctuators. We also include in our model two sinks which are represented by  quasi-degenerate manifolds of electron energy levels.  The sinks are directly coupled to the donor and acceptor states. This is done using the well-known Weisskopf-Wigner model \cite{WW,SM}.

When both noise and sinks influence the ET, the electron dynamics becomes rather complicated, and is generally described by a system of integro-differential  equations. We derive the conditions for which a simplified system of ordinary differential equations can be used.

Our approach is rigorous in the sense that all approximations are controlled and justified.  We obtain analytically and numerically the optimal ET rates  and efficiency for both sharp and flat redox potential. Our results are useful for analyzing and engineering optimal properties of photosynthetic bio-complexes.

\begin{acknowledgements}

This work was carried out under the auspices of the National Nuclear Security Administration of the U.S. Department of Energy at Los Alamos National Laboratory under Contract No. DE-AC52-06NA25396. RTS acknowledges from Office of Energy-Department of Energy-DE-SC0001295 for contributions regarding the organization of electron donors and acceptors in reaction center complexes.  A.I. Nesterov acknowledges the support from the CONACyT, Grant No. 15349.

\end{acknowledgements}
\appendix

\section{Non-Hermitian Effective Hamiltonian}

We consider the time-dependent Hamiltonian of a $N$-level system coupled with independent sinks through each level:
\begin{align}
&\mathcal H(t) = \sum^N_{n=1} \epsilon_n(t)|n\rangle\langle n| + \sum_{m \neq n} \beta_{nm}(t)|n\rangle\langle m| \nonumber \\
& +  \sum^{N}_{n=1}\sum^{N_n}_{i_n=1} ( E_{i_n}|i_n\rangle\langle  i_n | +V_{ni_n}|n\rangle\langle  i_n | + V_{i_n n}|i_n\rangle\langle n |\big),
\end{align}
where $m,n = 1,2, \dots,N$. We assume that the sinks are sufficiently dense, so that one can perform an integration instead of a summation. Then we have,
\begin{align}\label{ah1}
&\mathcal H(t) = \sum_{n}\epsilon_n(t)|n\rangle\langle n| + \sum_{m \neq n} \beta_{mn}(t)|n\rangle\langle m| \nonumber \\
&+\sum_{n} \Big( \int\alpha_n(E)|n\rangle\langle E|g_n(E) dE + \rm h.c.\Big) \nonumber \\
&+\sum_{n}\int E|E\rangle\langle E|g_n(E)dE
 \end{align}
where $g_n(E)$ is the density of states, and $V_{ni_n}\rightarrow \alpha_n(E)$.

With the state vector written as
\begin{align}\label{aS1}
|\psi\rangle = \sum_{n}\Big(c_n(t) |n\rangle + \int c_{nE}(t)|E\rangle g_n(E)dE \Big ),
\end{align}
the Schr\"odinger equation,
\begin{align}\label{ach10}
    i\frac{\partial |\psi(t) \rangle}{\partial t} = {\mathcal H} |\psi(t) \rangle,
\end{align}
takes the form
\begin{align}\label{ach11}
    i\dot c_n(t)=& E(t) c_n(t) + \sum_{m \neq n} \beta_{nm}(t) c_m(t) \nonumber \\
     &+ \int^\infty_{0}\alpha_n^\ast(E) c_{nE}(t)g_n(E) dE \\
    i\dot c_{nE}(t)=&E c_E(t) + \alpha_n(E) c_n(t) .
    \label{ach11a}
\end{align}

In order to eliminate the continuum amplitudes from the equations for the discrete states, we first apply the Laplace transformation:
\begin{align}\label{ach15}
c_n(t) = \int_0^\infty e^{-st} c_n(s) ds, \\
c_{nE}(t) = \int_0^\infty e^{-st} c_{nE}(s) ds.
\end{align}
Then, from Eq. (\ref{ach11a}) we obtain
\begin{align}\label{Ach16}
    (s + iE)c_{nE}(s) = - i  \alpha_n(E) c_{n}(s).
\end{align}
This yields $c_E(s) = - i \alpha_n(E) c_2(s)/(s + iE)$.
Inserting this expression for $c_E(s)$ into Eq. ({\ref{ach11}}), we obtain the following system of integro-differential equations, describing the non-Markovian dynamics of the TLS,
\begin{align}\label{ach17a}
    i\dot c_n(t)= &E c_n(t) + \sum_{n \neq m} \beta_{nm}(t) c_m(t) \nonumber \\
     &-i \int^{\infty}_{0}  c_{n}(s) e^{-st} ds\int \frac{|\alpha_n(E)|^2g_n(E) \, dE _n}{s + iE}.
\end{align}

To proceed further, we change the variable in the last integral, $s \rightarrow -E'$, so that
\begin{align}\label{Aeq1}
 \int \frac{|\alpha_n(E)|^2g_n(E) \, dE }{s + iE} \rightarrow  -i \int \frac{|\alpha_n(E)|^2g_n(E) \, dE _n}{E- E'}
\end{align}
The next step is to use the identity
\begin{align}\label{ach18}
    \frac{1}{x-x^\prime + i0} = {\mathcal P}  \bigg\{ \frac{1}{x-x^\prime }\bigg\} - i\pi\delta(x-x^\prime),
\end{align}
where ${\mathcal P}$ = Principal value. This yields
\begin{align}\label{ach17}
  \int \frac{|\alpha_n(E)|^2g_n(E) \, dE }{E- E'}=
  \Delta(E') -  \frac{i}{2}\Gamma_n(E'),
  \end{align}
where
\begin{align}\label{Aeq2}
 \Delta(E')  ={\mathcal P}  \int \frac{|\alpha_n(E)|^2g_n(E) \, dE }{E- E'} , \\
 \Gamma_n(E')=  2\pi   \int{|\alpha_n(E)|^2 }g_n(E){\delta(E-  E')}\, dE.
\end{align}

Now using the Weisskopf-Wigner pole approximation, we evaluate the integrals as follows \cite{WW,SM,KE}:
\begin{align}\label{ach19}
\Delta(E') \approx\Delta(\epsilon_n)&=\mathcal {P} \int \frac{|\alpha_n(E)|^2 g_n(E) dE }{E - \epsilon_n}, \\
 \Gamma_n(E')\approx \Gamma_n(\epsilon_n)& =
  2\pi\int{|\alpha_n(E)|^2 }g_n(E){\delta(E -\epsilon_n)}\, dE \nonumber \\
  &= 2\pi g_n(\epsilon_n)|\alpha_n(\epsilon_n)|^2 .
\end{align}
The Weisskopf-Wigner pole approximation basically corresponds to the assumption that the coupling constant to the continuum is a smoothly varying  function of the energy, e.g.
the continuum is treated as a single discrete level.

Inserting (\ref{ach19}) into Eq. (\ref{ach11}), we obtain
\begin{align}\label{eqach17}
    i\dot c_n(t)=  \varepsilon_n(t) c_n(t) + \sum_{m \neq n}  \beta_{nm}(t) c_m(t)  - \frac{i\Gamma_n}{2}  c_n(t),
\end{align}
where $\Gamma_n = \Gamma_n(E_n)$ and $\varepsilon(t) = \epsilon_n(t)  -\Delta(E_n)$.

Writing $|\psi_N\rangle = \sum_{n}c_n(t) |n\rangle $, we find that the dynamics of the $N$-level system interacting with the continuum is described by the Schr\"odinger equation,
\begin{align}\label{ach10a}
    i\frac{\partial |\psi_N(t) \rangle}{\partial t} =\tilde {\mathcal H} |\psi_N(t) \rangle,
\end{align}
where $ \tilde{\mathcal H}= {\mathcal H}- i \mathcal W$ is the effective non-Hermitian Hamiltonian,
 \begin{align}
 {\mathcal H} = \sum_{n}\varepsilon_n|n\rangle \langle n|  +\sum_{m \neq n}  \beta_{mn}(t) |m\rangle \langle n|
 \end{align}
being the dressed Hamiltonian, and
 \begin{align}
  W = \frac{1}{2}\sum_{n}\Gamma_n|n\rangle \langle n|.
 \end{align}

 Equivalently, the dynamics of this system can be described by the Liouville  equation,
 \begin{eqnarray}\label{DM1}
    i \dot{ \rho} =\tilde {\mathcal H}\rho - \rho{\tilde {\mathcal H}}^\dagger= [\mathcal H,\rho] - i\{\mathcal W,\rho\},
 \end{eqnarray}
 where   $\rho $ is the density matrix projected on the intrinsic states, and $\{\mathcal W,\rho\}= \mathcal W\rho +\rho\mathcal  W$.

 In particular case of the two-level system considered in this paper, the effective non-Hermitian Hamiltonian takes the form
 \begin{align}\label{ach12}
    \tilde {\mathcal H } =  \frac{1}{2} \left(
                                 \begin{array}{cc}
                                    2\varepsilon_1- i\Gamma_1 &  V \\
                                     V& 2\varepsilon_2- i\Gamma_2
                                   \end{array}
                               \right).
 \end{align}

{\em Comments.} The results of this section can be obtained using the standard Feshbach projection method \cite{RI,RI2,RI3,VZ,SM}.

\section{Equation of motion for the average density matrix}

In this Appendix, we derive from the Liouville equation,
$i\dot{ \rho} = [\tilde{\mathcal H},\rho] - i \{\mathcal W,\rho\}$, the equation of motion for the average density matrix. We will use the interaction representation. Considering the off-diagonal elements as perturbations, so that $\tilde{\mathcal H}={\mathcal H}_0 + V(t)- i\mathcal W$, where
$ {\mathcal H}_0= \sum_{n} \varepsilon_n |n\rangle\langle  n | +  \sum_{n} \lambda_{nn} (t) |n\rangle\langle  n | $, $V(t)= \sum_{m \neq n} ( V_{mn} +\lambda_{mn}(t))|m\rangle\langle  n |$, and  $\mathcal W =(\Gamma_1/2) |1\rangle \langle 1|+ (\Gamma_2/2) |2\rangle \langle 2|$, we obtain the following equations of motion:
\begin{align}  \label{A1a}
{\dot {\tilde \rho}}_{11} = i({\tilde \rho}_{12}{\tilde V}_{21}- {\tilde V}_{12} {\tilde \rho}_{21})- \Gamma_1 {\tilde \rho}_{11}, \\
{\dot {\tilde \rho}}_{22} = i({\tilde \rho}_{21}{\tilde V}_{12}- {\tilde V}_{21} {\tilde \rho}_{12})-\Gamma_2  {\tilde \rho}_{22},  \\
{\dot {\tilde \rho}}_{12} = i{\tilde V}_{12}({\tilde \rho}_{11}-  {\tilde \rho}_{22})- \Gamma  {\tilde \rho}_{12}, \\
{\dot {\tilde \rho}}_{21} = i{\tilde V}_{21}({\tilde \rho}_{11}-  {\tilde \rho}_{22}) -  \Gamma  {\tilde \rho}_{21},
\label{A1b}
\end{align}
where $\Gamma =(\Gamma_1 + \Gamma_2)/2$,
\begin{align}
\tilde \rho= T(e^{i\int_0^t H_0(\tau) d \tau})\rho T(e^{-i\int_0^t H_0(\tau) d\tau }),
\end{align}
and
\begin{align}
 \tilde V=  T(e^{i\int_0^t H_0(\tau) d \tau})V T(e^{-i\int_0^t H_0(\tau) d\tau }).
\end{align}
Using Eqs. (\ref{A1a}) - (\ref{A1b}), we obtain
\begin{widetext}
\begin{align} \label{Eq2}
 {\tilde \rho}_{11}(t) = &{ {\tilde \rho}}_{11}(0) + i\int_0^t e^{-\Gamma_1 (t- t')}({\tilde \rho}_{12}(t'){\tilde V}_{21}(t')- {\tilde V}_{12}(t') {\tilde \rho}_{21}(t'))dt', \\
 {\tilde \rho}_{22}(t) = &  {\tilde \rho}_{22}(0)+ i\int_0^t e^{-\Gamma_2 (t- t')}({\tilde \rho}_{21}(t'){\tilde V}_{12}(t')- {\tilde V}_{21}(t') {\tilde \rho}_{12}(t')) dt', \\
 {\tilde \rho}_{12}(t) =& {\tilde \rho}_{12}(0) + i \int_0^t e^{-\Gamma (t- t')}{\tilde V}_{12}(t') ({\tilde \rho}_{11}(t')-  {\tilde \rho}_{22}(t'))dt', \\
{\tilde \rho}_{21}(t) = &{\tilde \rho}_{21}(0) + i  \int_0^t e^{-\Gamma (t- t')}{\tilde V}_{21}(t')({\tilde \rho}_{11}(t')-  {\tilde \rho}_{22}(t'))dt'.
\label{Eq3a}
\end{align}
\end{widetext}

We assume that initially ${\tilde \rho}_{12}(0)={\tilde \rho}_{21}(0)=0$. Now, inserting  (\ref{Eq2}) - (\ref{Eq3a}) into Eqs. (\ref{A1a}) - (\ref{A1b}), and taking into account that ${\tilde \rho}_{11} =  \rho_{11}$ and ${\tilde \rho}_{22} =  \rho_{22}$,  we obtain the following system of integro-differential equations,
\begin{widetext}
\begin{align}  \label{B3}
{\dot {\rho}}_{11}(t) = & - \int_0^t e^{-\Gamma (t- t')}\Big({\tilde V}_{21}(t){\tilde V}_{12}(t')+ {\tilde V}_{21}(t'){\tilde V}_{12}(t)\Big)\Big({ \rho}_{11}(t') -{ \rho}_{22}(t')\Big) dt' - \Gamma_1 {\rho}_{11}(t), \\
{\dot {\rho}}_{22}(t) = & \int_0^te^{-\Gamma (t- t')}\Big({\tilde V}_{21}(t){\tilde V}_{12}(t')+ {\tilde V}_{21}(t'){\tilde V}_{12}(t)\Big)\Big({ \rho}_{11}(t') -{\rho}_{22}(t')\Big) dt' - \Gamma_2 {\rho}_{22}(t) , \\
\dot{\tilde \rho}_{12}(t) = & - \int_0^t\Big(e^{-\Gamma _1(t- t')}+ e^{-\Gamma _2(t- t')}\Big) \Big({\tilde V}_{21}(t'){\tilde \rho}_{12}(t')-  {\tilde V}_{12}(t'){\tilde \rho}_{21}(t')\Big ){\tilde V}_{12}(t)dt'- \Gamma  {\rho}_{12}(t) \nonumber \\
&+ i{\tilde V}_{12}(t)({ \rho}_{11}(0)-  { \rho}_{22}(0))  , \\
\dot{\tilde \rho}_{21}(t) =& - \int_0^t\Big(e^{-\Gamma _1(t- t')}+ e^{-\Gamma _2(t- t')}\Big) \Big({\tilde V}_{21}(t'){\tilde \rho}_{12}(t')-  {\tilde V}_{12}(t'){\tilde \rho}_{21}(t')\Big ){\tilde V}_{21}(t)dt'- \Gamma  {\rho}_{21}(t) \nonumber \\
&+ i{\tilde V}_{21}(t)({ \rho}_{11}(0)-  {\rho}_{22}(0)) .
 \label{B4}
\end{align}
\end{widetext}
For the average components of the density matrix this yields
\begin{widetext}
\begin{align}  \label{Eq3}
\frac{d}{dt}{\langle{\rho}}_{11}(t)\rangle  =& - \int_0^te^{-\Gamma (t- t')}\Big\langle\Big({\tilde V}_{21}(t){\tilde V}_{12}(t')+ {\tilde V}_{21}(t'){\tilde V}_{12}(t)\Big)\Big({ \rho}_{11}(t') -{ \rho}_{22}(t')\Big) \Big\rangle dt'  -\Gamma_1 \langle{\rho}_{11}(t)\rangle, \\
\frac{d}{dt}{\langle{\rho}}_{22}(t)\rangle  = & \int_0^te^{-\Gamma (t- t')}\Big\langle\Big({\tilde V}_{21}(t){\tilde V}_{12}(t')+ {\tilde V}_{21}(t'){\tilde V}_{12}(t)\Big)\Big({ \rho}_{11}(t') -{\rho}_{22}(t')\Big)\Big\rangle dt' - \Gamma_2 \langle{\rho}_{22}(t)\rangle , \\
\frac{d}{dt}{\langle{\rho}}_{12}(t)\rangle  = & - \int_0^t\Big(e^{-\Gamma _1(t- t')}+ e^{-\Gamma _2(t- t')}\Big)\Big\langle \Big({\tilde V}_{21}(t'){\tilde \rho}_{12}(t')-  {\tilde V}_{12}(t'){\tilde \rho}_{21}(t')\Big ){\tilde V}_{12}(t)\Big\rangle dt'- \Gamma  \langle{\rho}_{12}(t)\rangle \nonumber \\
&+ i\langle{\tilde V}_{12}(t)\rangle({ \rho}_{11}(0)-  { \rho}_{22}(0))  , \\
\frac{d}{dt}{\langle{\rho}}_{21}(t)\rangle  =& - \int_0^t\Big(e^{-\Gamma _1(t- t')}+ e^{-\Gamma _2(t- t')}\Big)\Big\langle \Big({\tilde V}_{21}(t'){\tilde \rho}_{12}(t')-  {\tilde V}_{12}(t'){\tilde \rho}_{21}(t')\Big ){\tilde V}_{21}(t)\Big\rangle dt'- \Gamma \langle {\rho}_{21}(t)\rangle \nonumber \\
&+ i\langle{\tilde V}_{21}(t)\rangle({\rho}_{11}(0)-  {\rho}_{22}(0)) ,
 \label{Eq4}
\end{align}
\end{widetext}
where the average $\langle\; \rangle$ is taken over the random process describing noise.

In the spin-fluctuator model of noise with the number of fluctuators, ${\mathcal N} \gg 1$, one has the following relations for the splitting of correlations \cite{NB1b},
\begin{align}
\big\langle\big({\tilde V}_{21}(t){\tilde V}_{12}(t')+ {\tilde V}_{21}(t'){\tilde V}_{12}(t)\big)\big({\tilde\rho}_{11}(t') -{\tilde\rho}_{22}(t')\big)\big \rangle = \nonumber \\
\big(\big\langle{\tilde V}_{21}(t){\tilde V}_{12}(t') \big\rangle + \big\langle{\tilde V}_{21}(t'){\tilde V}_{12}(t)\big\rangle\big)\big(\big\langle{\tilde\rho}_{11}(t')\big\rangle -\big\langle{\tilde\rho}_{22}(t')\big\rangle\big),
\end{align}
and so on. Using these relations, we obtain the following system of integro-differential equations for the average components of the density matrix,
\begin{widetext}
\begin{align} \label{Eq8a}
\frac{d}{dt}{\langle{\rho}}_{11}(t)\rangle =&-\int_0^t e^{-\Gamma (t- t')}\big(\big\langle{\tilde V}_{21}(t){\tilde V}_{12}(t') \big\rangle + \big\langle{\tilde V}_{21}(t'){\tilde V}_{12}(t)\big\rangle\big)\big(\big\langle{\rho}_{11}(t')\big\rangle -\big\langle{\rho}_{22}(t')\big\rangle\big)dt'- \Gamma_1 \langle { \rho}_{11}(t)\rangle,  \\
\frac{d}{dt}{\langle{\rho}}_{22}(t)\rangle =&\int_0^t e^{-\Gamma (t- t')}\big(\big\langle{\tilde V}_{21}(t){\tilde V}_{12}(t') \big\rangle + \big\langle{\tilde V}_{21}(t'){\tilde V}_{12}(t)\big\rangle\big)\big(\big\langle{\rho}_{11}(t')\big\rangle -\big\langle{\rho}_{22}(t')\big\rangle\big)dt' - \Gamma_2 \langle { \rho}_{22}(t)\rangle, \\
\frac{d}{dt}\langle {\tilde \rho}_{12}(t)\rangle = & i\langle{\tilde V}_{12}(t)\rangle({ \rho}_{11}(0)-  { \rho}_{22}(0)) - \int_0^t \Big(e^{-\Gamma _1(t- t')}+ e^{-\Gamma _2(t- t')}\Big) \langle{\tilde V}_{12}(t) {\tilde V}_{21}(t')\rangle \langle{\tilde \rho}_{12}(t')\rangle dt' \nonumber \\
&+  \int_0^t \Big(e^{-\Gamma _1(t- t')}+ e^{-\Gamma _2(t- t')}\Big) \langle{\tilde V}_{12}(t) {\tilde V}_{12}(t')\rangle \langle{\tilde \rho}_{21}(t')\rangle dt'  - \Gamma  \langle{\tilde \rho}_{12}(t)\rangle,\\
\frac{d}{dt}\langle {\tilde \rho}_{21}(t)\rangle = & i\langle{\tilde V}_{21}(t)\rangle({ \rho}_{11}(0)-  { \rho}_{22}(0))
- \int_0^t \Big(e^{-\Gamma _1(t- t')}+ e^{-\Gamma _2(t- t')}\Big) \langle{\tilde V}_{21}(t) {\tilde V}_{21}(t')\rangle \langle{\tilde \rho}_{12}(t')\rangle dt'  \nonumber \\
& +  \int_0^t \Big(e^{-\Gamma _1(t- t')}+ e^{-\Gamma _2(t- t')}\Big)\langle{\tilde V}_{21}(t) {\tilde V}_{12}(t')\rangle \langle{\tilde \rho}_{21}(t')\rangle dt'  - \Gamma  \langle{\tilde\rho}_{21}(t)\rangle .
\end{align}
\end{widetext}

Using these relations, we obtain the following system of integro-differential equations for the diagonal components of the density matrix,
\begin{widetext}
\begin{align} \label{B4ar}
\frac{d}{dt}{\langle{\rho}}_{11}(t)\rangle =&- \int_0^t { K}(t,t')\big(\big\langle{\rho}_{11}(t')\big\rangle -\big\langle{\rho}_{22}(t')\big\rangle\big)dt' - \Gamma_1 \langle { \rho}_{11}(t)\rangle, \\
\frac{d}{dt}{\langle{\rho}}_{22}(t)\rangle =&\int_0^t { K}(t,t')\big(\big\langle{\rho}_{11}(t')\big\rangle -\big\langle{\rho}_{22}(t')\big\rangle\big)dt' - \Gamma_2 \langle { \rho}_{22}(t)\rangle,
\label{B4br}
\end{align}
\end{widetext}
where the kernel, $K(t,t')$, is given by
\begin{align}\label{AK1}
  K(t,t') = e^{-\Gamma (t- t')}\big(\big\langle{\tilde V}_{21}(t){\tilde V}_{12}(t') \big\rangle + \big\langle{\tilde V}_{21}(t'){\tilde V}_{12}(t)\big\rangle\big).
\end{align}

For the diagonal noise, so that $\lambda_{mn}=0, \quad (m\neq n)$, the kernel can be recast as
 \begin{align}\label{AK2}
  K(t -t') =\frac{V^2}{2} \cos(\varepsilon(t-t')) e^{-\Gamma (t- t')} \big\langle e^{i\kappa(t-t')}\big \rangle,
\end{align}
where $\kappa(t-t') = D\int_0^{t-t}\xi(\tau)d\tau$ and $D =|g_1-g_2|$.

In the Gaussian approximation, the generating functional becomes
\begin{align}
\big\langle e^{i\kappa(t-t')} \big\rangle =  e^{-\langle\kappa^2(t-t')\rangle/2}.
\label{AEq15a}
\end{align}
\begin{align}
\langle\kappa^2(t-t')\rangle=2 D^2\int^{t-t'}_0 d\tau'\int_{0}^{\tau'}d\tau''\chi(\tau' -\tau'').
\label{G2}
\end{align}

With help of Eqs. (\ref{AEq15a}) - (\ref{G2}) we obtain
\begin{widetext}
\begin{align}
K(t -t') =\frac{V^2}{2} \cos(\varepsilon(t-t'))\exp\bigg(-\Gamma(t-t')- D^2\int^{t-t'}_0 d\tau'\int_{0}^{\tau'}d\tau''\chi(\tau' -\tau'') \bigg).
\label{AEq15b}
\end{align}
\end{widetext}

Using the results obtained in Sec. III,  one can show that for $V< D\sigma$ the system of integro-differential equations (\ref{B4ar}) - (\ref{B4br}) can be approximated by the following system of ordinary differential equations:
\begin{align} \label{BN6ar}
\frac{d}{dt}{\langle{\rho}}_{11}\rangle =&- {\mathfrak R}(t)\big(\big\langle{\rho}_{11}\big\rangle -\big\langle{\rho}_{22}\big\rangle\big) -  \Gamma_1 \langle {\rho}_{11}\rangle, \\
\frac{d}{dt}{\langle{\rho}}_{22}\rangle = &\,{\mathfrak R}(t)\big(\big\langle{\rho}_{11}\big\rangle -\big\langle{\rho}_{22}\big\rangle\big)  - \Gamma_2 \langle {\rho}_{22}\rangle  ,
\label{BN7r}
\end{align}
where ${\mathfrak R}(t)= \int_{0}^{t} \tau K(\tau) d\tau $. Performing the integration we obtain
\begin{align}
{\mathfrak R}(t)=& \frac{\sqrt{\pi}q}{4p}\exp\bigg(\frac{q^2}{4p^2}\bigg)\bigg({\rm erf}\bigg(\frac{q}{2p} +pt\bigg )- {\rm erf}\bigg(\frac{q}{2p}\bigg)\bigg )\nonumber \\
  &+\frac{\sqrt{\pi}\bar q}{4p} \exp\bigg(\frac{{\bar q}^2}{4p^2}\bigg)\bigg({\rm erf}\bigg(\frac{\bar q}{2p}+pt\bigg)- {\rm erf}\bigg(\frac{\bar q}{2p}\bigg)\bigg),
\label{R1}
\end{align}
where $p=D\sigma/\sqrt{2}$, $q=\Gamma+ i\varepsilon$, $\bar q=\Gamma-  i\varepsilon$, and ${\rm erf}(z)$ is the error function \cite{abr}.


\end{document}